\documentclass[journal]{IEEEtran}

\usepackage{amssymb}
\usepackage[dvips]{graphicx}
\usepackage[usenames]{color}
\usepackage{amsfonts}
\usepackage{bm}
\usepackage{subfigure}
\usepackage{booktabs,multirow}
\usepackage{color}
\usepackage{setspace}
\usepackage{algorithm} 
\usepackage{algorithmic} 

\usepackage{cite}
\usepackage[cmex10]{amsmath}

\IEEEoverridecommandlockouts

   \renewenvironment{thebibliography}[1]{%
     \begin{oldthebibliography}{#1}%
       \setlength{\parskip}{0ex}%
       \setlength{\itemsep}{0ex}%
   }%
   {%
     \end{oldthebibliography}%
   }

\begin{document}
\title{\mbox{}\vspace{0.40cm}\\
\textsc{Joint Optimization of Computation and Communication Power in Multi-user Massive MIMO Systems} \vspace{0.2cm}}

\author{\normalsize
Xiaohu~Ge,~\IEEEmembership{Senior Member,~IEEE,}
        Yang~Sun,
        Hamid~Gharavi,~\IEEEmembership{Life Fellow,~IEEE,}
        and~John~Thompson,~\IEEEmembership{Fellow,~IEEE,}

\thanks {Xiaohu Ge, Yang Sun are with School of Electronic Information and Communications, Huazhong University of Science and Technology, Wuhan 430074, Hubei, P. R.
China. Email: \{xhge, yang$\_$sun\}@mail.hust.edu.cn.}
\thanks {Hamid Gharavi is with National Institute of Standards and Technology (NIST),Gaithersburg, MD 20899-8920 USA. Email: hamid.gharavi@nist.gov.}
\thanks {John Thompson is with Institute for Digital Communications, University of Edinburgh, Edinburgh, EH9 3JL, UK. Email: john.thompson@ed.ac.uk.}
}
\maketitle
\begin{abstract}
With the growing interest in the deployment of massive multiple-input-multiple-output (MIMO) systems and millimeter wave technology for fifth generation (5G) wireless systems, the computation power to the total power consumption ratio is expected to increase rapidly due to high data traffic processing at the baseband unit. Therefore in this paper, a joint optimization problem of computation and communication power is formulated for multi-user massive MIMO systems with partially-connected structures of radio frequency (RF) transmission systems. When the computation power is considered for massiv MIMO systems, the results of this paper reveal that the energy efficiency of massive MIMO systems decreases with increasing the number of antennas and RF chains, which is contrary with the conventional energy efficiency analysis results of massive MIMO systems, i.e., only communication power is considered. To optimize the energy efficiency of multi-user massive MIMO systems, an upper bound on energy efficiency is derived. Considering the constraints on partially-connected structures, a suboptimal solution consisting of baseband and RF precoding matrices is proposed to approach the upper bound on energy efficiency of multi-user massive MIMO systems. Furthermore, an oPtimized Hybrid precOding with computation and commuNication powEr (PHONE) algorithm is developed to realize the joint optimization of computation and communication power. Simulation results indicate that the proposed algorithm improves energy and cost efficiencies and the maximum power saving is achieved by 76.59\% for multi-user massive MIMO systems with partially-connected structures.
\end{abstract}


\section{Introduction}
\label{sec1}
To meet the anticipated high volume of traffic demand for fifth generation (5G) wireless communication systems, massive multiple-input-multiple-output (MIMO) and millimeter wave technologies are emerging as key solutions \cite{Rangan16,Akdeniz14}. However, it is impractical to perform a fully digital precoding solution, \emph{i.e.}, zero-forcing, for massive MIMO systems and millimeter wave technologies due to power consumption and space constraints in the analog front-end \cite{Heath16}. To reduce communication power consumption and the number of radio frequency (RF) chains, a hybrid analogue/digital precoding is proposed as a viable approach for the deployment of massive MIMO systems with millimeter wave technology \cite{Han15,Zhang15}.  Moreover, this technology is expected to have a major impact in promoting small cells as the main cellular architecture in 5G wireless networks. Bear in mind that, unlike traditional macro cells, the computation power of small cells equipped with massive MIMO systems can consume more than 40\% of  the total power \cite{Auer10,Ge17}. Therefore, our main objective in this paper is to improve the energy efficiency by jointly optimizing the computation and communication power for multi-user massive MIMO systems.

Generally, there exist two types of hybrid precoding solutions in RF systems: the fully-connected structure and the partially-connected structure \cite{Heath16}. In the former, all the antennas are connected to each RF chain by phase shifters where multiplexing between RF chains and antennas can be achieved for massive MIMO systems \cite{Ayach13,Rusu16,Chen15,Alkhateeb14,Liang14,Ni15,Sohrabi16,Zi16}. For instance, a hybrid precoding solution with fully-connected structures using orthogonal matching pursuit that utilizes the structure of millimeter wave channels was proposed in \cite{Ayach13}. Based on the simulation results, it was observed that the proposed algorithm can approach the theoretical limit of spectral efficiency. As a trade-off between performance and complexity, four precoding hybrid algorithms were investigated in \cite{Rusu16} for a single user massive MIMO system. An algorithm based on iteratively updating phases of phase shifters in the RF precoder was proposed in [10]. The proposed approach aims at minimizing the weighted sum of squared residuals between the optimal full-baseband design and the hybrid design.  To guarantee that precoding can converge to a locally optimal solution, a hybrid precoding algorithm was developed in \cite{Chen15}.  In addition, a hybrid precoding algorithm based on an adaptive channel estimation was developed in \cite{Alkhateeb14}. The proposed algorithm aims at relaxing hardware constraints on the analogue only beamforming and achieving spectral efficiency of fully digital solutions \cite{Alkhateeb14}.  Considering multi-user massive MIMO scenarios, a hybrid precoding scheme that approaches spectral efficiency of a  traditional baseband zero-forcing (ZF) precoding scheme was proposed in \cite{Liang14}.  Furthermore, to harvest a large array gain through phase-only RF precoding, a hybrid block diagonalization (BD) scheme capable of approaching the capacity of the traditional BD processing method in massive MIMO systems was investigated in \cite{Ni15}. When the number of RF chains is less than twice the number of data streams, the authors in \cite{Sohrabi16} developed a heuristic algorithm to solve the problem of spectral efficiency maximization for transmission scenarios, such as a point-to-point massive MIMO system and a multi-user multiple-input-single-output (MISO) system. Based on the fully-connected structure, \cite{Zi16} developed
a hybrid precoding scheme to optimize the energy efficiency of multi-user massive MIMO systems.

Although the fully-connected structure of a hybrid precoding solution can approach the theoretical limit of spectral efficiency for fully digital precoding systems, the partially- connected hybrid precoding approach (\emph{i.e.}, every RF chain is connected to a limited number of antennas) is more attractive for practical implementation due to low complexity and cost \cite{Heath16, Roh14,Kim13,Zhou16,Yu16,Lin16,He17,Gao16}.  A comparison between fully-connected and partially-connected structures of hybrid precoding for massive MIMO systems with millimeter wave technology was performed in \cite{Heath16}, which indicates that the partially-connected structure of a hybrid precoding solution can offer a potential advantage of balancing cost and performance for massive MIMO systems. Furthermore, based on a prototype system, the advantages of a hybrid beamforming scheme in 5G cellular networks with a partially-connected structure were demonstrated \cite{Roh14}. In \cite{Kim13} a multi-beam transmission diversity scheme was proposed for single stream and single user case in massive MIMO systems with partially-connected structures \cite{Kim13}. To improve the transmission rate, a hybrid precoding scheme for partially-connected structures capable of adaptively adjusting the number of data streams was developed by \cite{Zhou16}. This approach is based on the rank of an equivalent baseband MIMO channel matrix and the received signal- to-noise ratio (SNR). Treating the hybrid precoder design as a matrix factorization problem, \cite{Yu16} proposes effective alternating minimization algorithms that can be used to optimize the transmission rate of massive MIMO systems with partially-connected structures. Considering the issue of power consumption in massive MIMO systems, energy efficiency optimization of a hybrid precoding solution with a partially-connected structure was studied in \cite{Lin16,He17,Gao16}. For instance, for multi-user massive MIMO systems with millimeter wave technology, it was shown that the partially-connected structure can outperform the fully-connected structure in terms of both spectral efficiency and energy efficiency \cite{Lin16}. Considering a single user massive MIMO system, the baseband and RF precoding matrices were optimized to improve energy efficiency of massive MIMO systems \cite{He17}. Based on the successive interference cancellation (SIC)-based hybrid precoding method, the authors in \cite{Gao16} have shown that energy efficiency of a single user massive MIMO system can be improved with low complexity.

It is partially-connected structure that attracts practical implementation. However, researches on it are rare, especially on energy efficiency optimization. Moreover, all the aforementioned studies which optimize energy efficiency for partially-connected structures use simple precoding optimization methods, such as optimizing baseband and RF precoding independently. Although the ratio of computation power to total power has shown improvement in massive MIMO systems, detailed investigation of the computation power model used for massive MIMO systems has received little attention in the open literature. In fact, these investigations simply treat energy consumption of massive MIMO systems solely as communication power \cite{Lin16,He17,Gao16}.

Motivated by the above gaps, in this paper we derive a joint optimization of computation and communication power for multi-user massive MIMO systems with partially-connected structures. The contributions and novelties of this paper are summarized as follows.
\begin{enumerate}
\item Considering that computation power consumes more than 40\% of the total power in massive MIMO systems, a new power consumption model that includes computation and communication power, is proposed to optimize the energy efficiency of massive MIMO systems.
\item Considering the joint optimization of computation and communication power, a new energy efficient optimization model is proposed for multi-user massive MIMO systems, which is based on partially-connected structures. The upper bound of energy efficiency is derived for multi-user massive MIMO systems with partially-connected structures. Then, utilizing the alternating minimization method, a suboptimal solution is derived for the baseband and RF precoding matrices to optimize energy efficiency. In contrast to the conventional energy efficiency optimization, \emph{i.e.}, focusing on communication power optimization in MIMO systems, the proposed energy efficiency suboptimal solution can jointly improve computation and communication power in massive MIMO systems.
\item Previous studies reveal that the energy efficiency of massive MIMO systems improves by increasing the numbers of antennas and RF chains when only the communication power is considered. However, our simulation results indicate that the energy efficiency of massive MIMO systems decreases with an increasing number of antennas and RF chains when computation and communication powers are considered. Moreover, simulation results show that the proposed algorithm for partially-connected structures outperforms that of fully-connected structures in energy and cost efficiency of multi-user massive MIMO systems. For example, when RF chains number is fixed at 14, the maximum power saving is achieved at 76.59\% and 38.38\% for multi-user massive MIMO communication systems with partially-connected and fully-connected structures, respectively.

\end{enumerate}

The remainder of this paper is organized as follows. Section~\ref{sec2} describes the system model of multi-user massive MIMO systems. In Section~\ref{sec3}, the energy and cost efficiencies are formulated for multi-user massive MIMO systems by adopting partially-connected structures. Section~\ref{sec4} presents the proposed hybrid precoding optimization solution for multi-user massive MIMO systems based on partially-connected structures. Simulation results and analysis are presented in Section~\ref{sec5}. Finally, conclusions are drawn in Section~\ref{sec6}.


\begin{figure}
\vspace{0.1in}
\centerline{\includegraphics[width=8cm,draft=faulse]{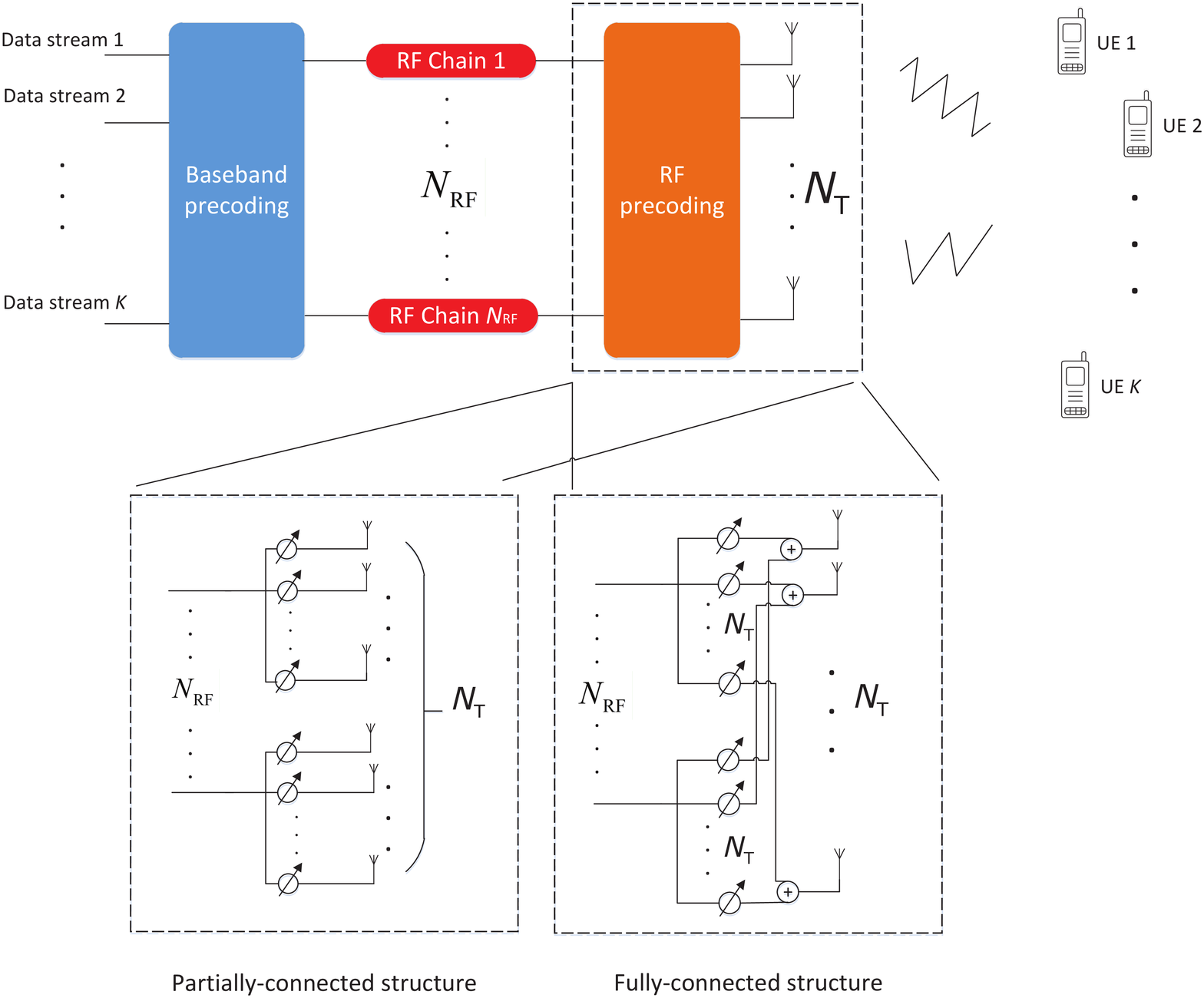}}
{\small Fig. 1. \@ \@ Multi-user massive MIMO communication systems.}
\end{figure}

\section{System Model}
\label{sec2}
Although the fully-connected structure of massive MIMO RF systems can easily approach the spectral efficiency limit for multi-user massive MIMO systems, the cost and complexity of massive MIMO RF systems are becoming a major issue for their future deployment. To reduce cost and simplify complexity of massive MIMO RF systems, the partially-connected structure, where each RF chain corresponds to multiple phase shifters and antennas as shown in Fig. 1, is a promising solution for industrial applications. In this paper, we jointly optimize the computation and communication power of massive MIMO RF systems to reduce the cost and complexity of multi-user massive MIMO systems with the partially-connected structure.

\subsection{Wireless Transmission Model}
\label{sec2-1}
A multi-user massive MIMO communication system with the partially-connected or fully-connected structures is illustrated in Fig. 1. As can be observed the transmitter in the massive MIMO communication system includes: a baseband unit with $K$  input data streams,  ${N_{{\rm{RF}}}}$ RF chains, and ${N_{\rm{T}}} \ge 100$  antennas. Considering the partially-connected structure, one RF chain is connected with $\frac{{{N_{\rm{T}}}}}{{{N_{{\rm{RF}}}}}}$  phase shifters and $\frac{{{N_{\rm{T}}}}}{{{N_{{\rm{RF}}}}}}$ antennas in such a way that antennas connected to each RF chain do not overlap. The receivers are configured as $K$ active user equipment (UEs), each with a single antenna. In this paper we focus on the downlink of multi-user massive MIMO communication systems.

The received signal at the $k{\rm{th}}$  UE is expressed by
\[{y_{k}} = {\bf{h}}^H_{k}{{\bf{B}}_{{\rm{RF}}}}{{\bf{B}}_{{\rm{BB}}}}{\bf{x}} + {w_{k}}.\tag{1}\]
In the above, ${\bf{x}} = {\left[ {{x_1},...,{x_{k}},...,{x_{K}}} \right]^H}$ is the signal vector transmitted from the transmitter to $K$ UEs, where ${x_{k}}$ is assumed to be independently and identically distributed (\emph{i.i.d.}) Gaussian random variables with zero mean and a variance of 1,  ${\bf{h}}_{k}$ is the downlink channel vector between the BS and the $k{\rm{th}}$  UE, and ${w_{k}}$ is the noise received by the $k{\rm{th}}$  UE. Moreover, all noise samples in the UE are \emph{i.i.d.} Gaussian random variables with zero mean and variance of $\sigma _{\rm{n}}^2$. The ${{\bf{B}}_{{\rm{BB}}}} \in {\mathbb{C}^{{N_{{\rm{RF}}}} \times K}}$ is the baseband precoding matrix, where the $k{\rm{th}}$  column of ${{\bf{B}}_{{\rm{BB}}}}$ is denoted as ${{\bf{b}}_{{\rm{BB}},{k}}}$  which is the baseband precoding vector for the $k{\rm{th}}$  UE. The ${{\bf{B}}_{{\rm{RF}}}} \in {\mathbb{C}^{{N_{\rm{T}}} \times {N_{{\rm{RF}}}}}}$ is the RF precoding matrix, which is realized by ${N_{\rm{T}}}$ phase shifters. For the partially-connected structure, every RF chain is equipped with an antenna sub-array as shown in Fig. 1. In this case, the RF precoding matrix ${{\bf{B}}_{{\rm{RF}}}}$ is a block diagonal matrix, \emph{i.e.}, ${{\bf{B}}_{{\rm{RF}}}} = {\rm{diag}}\{ {{\bf{m}}_{\rm{1}}},...,{{\bf{m}}_{i}}, \ldots ,{{\bf{m}}_{{N_{{\rm{RF}}}}}}\} $ , where ${{\bf{m}}_{i}} $ is the  $i{\rm{th}}$ block matrix which corresponds to the precoding matrix between the $i{\rm{th}}$ RF chain and the connected $\frac{{{N_{\rm{T}}}}}{{{N_{{\rm{RF}}}}}}$ antennas, ${{\bf{m}}_{i}} $ is a $\frac{{{N_{\rm{T}}}}}{{{N_{{\rm{RF}}}}}} \times 1$ complex vector and the amplitude of vector element is fixed as 1. When the bandwidth of the $k{\rm{th}}$  UE is configured as $W$, considering interference caused by sidelobe beam, the available rate of the $k{\rm{th}}$ UE is expressed by
\[{R_{k}} = W{\log _2}\left(1 + \frac{{{\bf{h}}^H_{k}{{\bf{B}}_{{\rm{RF}}}}{{\bf{b}}_{{\rm{BB}},{k}}}{\bf{b}}_{{\rm{BB}},{k}}^H{\bf{B}}_{{\rm{RF}}}^H{\bf{h}}_{k}}}{{\sum\limits_{i = 1,i \ne k}^K {{\bf{h}}^H_{k}{{\bf{B}}_{{\rm{RF}}}}{{\bf{b}}_{{\rm{BB}},{i}}}{\bf{b}}_{{\rm{BB}},{i}}^H{\bf{B}}_{{\rm{RF}}}^H{\bf{h}}_{k}}  + \sigma _{\rm{n}}^2}}\right),\tag{2}\]
where superscript $H$ is the conjugate transposition operation on the matrix.

When the transmissions of all UEs are considered, the available sum rate of multi-user massive MIMO communication system is expressed by
\[{R_{{\rm{sum}}}} = \sum\limits_{{k} = 1}^K {{R_{k}}}.\tag{3} \]

To support massive wireless traffic in 5G wireless communication systems, millimeter wave technology is adopted for multi-user massive MIMO communication systems. Based on the propagation characteristic of millimeter wave in wireless communications, a geometry-based stochastic model (GBSM) is used to describe the millimeter wave channel of multi-user massive MIMO communication systems \cite{Xu02,Sayeed02,Bogale14}
\[{{\bf{h}}_{k}} = \sqrt {\frac{{{N_{\rm{T}}}{\varepsilon _{k}}}}{{{N_{{\rm{ray}}}}}}} \sum\limits_{i = 1}^{{N_{{\rm{ray}}}}} {{\rho _{{ki}}}{{\bf{u}}}\left( {{\theta _{i}},{\vartheta _{i}}} \right)},\tag{4} \]
where ${N_{{\rm{ray}}}}$ is the number of the multi-paths between the transmitter and $K$  UEs, ${\varepsilon _{k}}$ is the path loss between the transmitter and the $k{\rm{th}}$  UE, ${\rho _{{ki}}}$ is the complex gain of the $k{\rm{th}}$  UE over the $i{\rm{th}}$  multi-path,  ${\theta _{i}}$ and ${\vartheta _{i}}$  are the azimuth and elevation angle of the $i{\rm{th}}$  multi-path over the antenna array at the transmitter, respectively. The ${\bf{u}}\left( {{\theta _{i}},{\vartheta _{i}}} \right)$  is the response vector of transmitter antenna array with the azimuth ${\theta _{i}}$ and elevation angle ${\vartheta _{i}}$. By assuming a uniform planar antenna array for the sake of simplicity, the response vector of transmitter antenna array with the azimuth ${\theta _{i}}$  and elevation angle ${\vartheta _{i}}$  is expressed as \cite{Bjrnson15}

\[\begin{array}{l}
{\bf{u}}\left( {{\theta _{i}},{\vartheta _{i}}} \right) = \frac{1}{{\sqrt {{N_{\rm{T}}}} }} [ {1,.} ..,\exp j\frac{{2\pi }}{\lambda }d( m \rm{sin} \left( {{\theta _{i}}} \right)\rm{sin}\left( {{\vartheta _{i}}} \right) + \\
\emph{n}\rm{cos}\left( {{\vartheta _{i}}} \right) ),
..., \exp j\left( {{N_{\rm{T}}} - 1} \right)\frac{{2\pi }}{\lambda }d(\left( {M - 1} \right)\rm{sin} \left( {{\theta _{i}}} \right)\rm{sin}\left( {{\vartheta _{i}}} \right) \\
+\left( {\emph{N} - 1} \right)\rm{cos}\left( {{\vartheta _{i}}} \right)) ]^T,
\end{array}\tag{5}\]

where $d$  is the distance between adjacent antennas, $\lambda $  is the carrier wavelength, $M$  and $N$  are the number of rows and columns of the transmitter antenna array, respectively. $m$ and $n$ represent the $m{\rm{th}}$ and $n{\rm{th}}$ antenna corresponding to the transmitter antenna array $(0 \le m < M,0 \le n < N)$, and $T$ is the transposition operation over the vector.

\subsection{Power Model }
\label{sec2-2}
Since the massive traffic data needs to be computed at the baseband unit and RF transmission systems, the computation power cannot be ignored for multi-user massive MIMO communication systems. Based on results in \cite{Desset12,Auer11,Tang12}, we express the total power at the transmitter as
\[{P_{{\rm{total}}}} = {P_{{\rm{commun}}}} + {P_{{\rm{CMPT}}}} + {P_{{\rm{fix}}}},\tag{6}\]
where ${P_{{\rm{commun}}}}$ is the communication power, ${P_{{\rm{CMPT}}}}$  is the computation power and  ${P_{{\rm{fix}}}}$ is the fixed power at the transmitter of multi-user massive MIMO communication systems. In general, the fixed power ${P_{{\rm{fix}}}}$ includes the cooling power, losses incurred by direct-current to direct-current (DC-DC) power supply and the mains power supply.

The communication power of multi-user massive MIMO communication systems is consumed by the power amplifiers (PAs) and RF chains, which is extended by
\[{P_{{\rm{commun}}}} = {P_{{\rm{PA}}}} + {P_{{\rm{RF}}}},\tag{7}\]
where ${P_{{\rm{PA}}}}$  is the power consumed by PAs and is calculated by
\[{P_{{\rm{PA}}}} = \frac{1}{\alpha }\sum\limits_{k = 1}^K {{{\left\| {{{\bf{B}}_{{\rm{RF}}}}{{\bf{b}}_{{\rm{BB,}\emph{k}}}}} \right\|}_{\rm{F}}^2}},\tag{8}\]
where $\alpha $ is the efficiency factor of PAs,${\left\| {} \right\|_{\rm{F}}}$ represent the Frobenius-norm. The power consumed at RF chains is expressed by \cite{Liu13}
\[{P_{{\rm{RF}}}} = {N_{{\rm{RF}}}}{P_{{\rm{one\_RF}}}},\tag{9}\]
where ${P_{{\rm{one\_RF}}}}$  is the power consumed of an RF chain. Substitute (8) and (9) into (7), the communication power of multi-user massive MIMO communication systems is expressed by
\[{P_{{\rm{commun}}}} = \frac{1}{\alpha }\sum\limits_{k = 1}^K {{{\left\| {{{\bf{B}}_{{\rm{RF}}}}{{\bf{b}}_{{\rm{BB}},{k}}}} \right\|}^2}}  + {N_{{\rm{RF}}}}{P_{{\rm{one\_RF}}}}.\tag{10}\]
The computation power of multi-user massive MIMO communication systems is consumed by wireless channel estimation, channel coding, linear processing at the baseband units and RF transmission systems, and processing to derive the precoding matrix, which is expressed by \cite{Bjrnson15}
\[{P_{{\rm{CMPT}}}} = {P_{{\rm{CE}}}} + {P_{{\rm{CD}}}} + {P_{{\rm{LP}}}} + P_{\rm{complex}},\tag{11}\]
where  ${P_{{\rm{CE}}}}$ is the power consumed by wireless channel estimation, ${P_{{\rm{CD}}}}$  is the power consumed by channel coding and ${P_{{\rm{LP}}}}$  is the power consumed by linear processing at baseband units and RF transmission systems, $P_{\rm{complex}}$ is the power consumed by our proposed algorithm to generate the precoding matrix.

To avoid explicit estimation of the channel, in this paper, channel estimation is done by beam training. For simplicity, the estimation power is obtained as a product of the number of subcarriers ($N$) times the number of paths of each subcarrier ($N_{ray}$), times  the estimated power of a subcarrier in a single path. For the latter term, the channel is assumed to be frequency -flat  and can be expressed as ${\kappa ^2}\frac{2}{{\overline \gamma  }}\left( {\frac{{({\kappa ^2} - 1){{\log }_\kappa }N}}{\delta } - 2} \right)\sum\limits_{s = 1}^{{{\log }_\kappa }N} {\frac{1}{{{G^{{\rm{BS}}}}(s)}}}$\cite{Alkhateeb14}, where $\kappa $ is the number of BS precoding vectors used in each training stage. $N$ is the number of discrete points taken from the angle of departure (AOD) quantization, $\overline \gamma$ is the average channel SNR, $\delta$ is the probability of estimation error. ${G^{{\rm{BS}}}}(s) = {N_{\rm{T}}}C_s^2$ is the beamforming gain at stage s, $C_s$ is a normalization constant.
So the channel estimation power for OFDM systems is derived as
\[{P_{{\rm{CE}}}} = K{N_{{\rm{ray}}}}{\kappa ^2}\frac{2}{{\overline \gamma  }}\left( {\frac{{({\kappa ^2} - 1){{\log }_\kappa }N}}{\delta } - 2} \right)\sum\limits_{s = 1}^{{{\log }_\kappa }N} {\frac{1}{{{G^{{\rm{BS}}}}(s)}}}.\tag{12} \]

Without loss of generality, the power of channel coding is assumed to be proportional to the available sum rate of a multi-user massive MIMO communication system \cite{Bjrnson15}. Therefore, the power of channel coding is expressed by
\[{P_{{\rm{CD}}}} = {P_{{\rm{COD}}}}\sum\limits_{k = 1}^K {{R_{k}}} ,\tag{13}\]
where  ${P_{{\rm{COD}}}}$ is the efficiency factor of channel coding, \emph{i.e.}, measured in Watt per bit per second.

We assume that the power of linear processing in multi-user massive MIMO communication systems is limited to the power consumed for precoding at both baseband units and RF transmission systems. Under these conditions, the power of linear processing can be extended as
\[{P_{{\rm{LP}}}} = {P_{{\rm{LP\_BB}}}} + {P_{{\rm{LP\_RF}}}},\tag{14}\]
where ${P_{{\rm{LP\_BB}}}}$  is the power consumed for the precoding at the baseband units and ${P_{{\rm{LP\_RF}}}}$ is the power consumed for the precoding within the RF transmission systems. Regardless of the Channel State Information (CSI) or precoding algorithm, the former, which is caused by the product of the signal vector times precoding matrix, can be expressed as: ${P_{{\rm{LP\_BB}}}} = \frac{\varphi {\nu _{{\rm{PCD}}}} }{{{L_{{\rm{TR}}}}}}$, where $\varphi $ is the number of floating-point computations in one baseband precoding operation, ${\nu _{{\rm{PCD}}}}$ is the number of baseband precodings per second, and ${L_{{\rm{TR}}}}$ is the computation efficiency of the transmitter. In this paper one baseband precoding operation is assumed to handle K symbols. Moreover, one symbol is configured to contain $\ell $ bits. To satisfy the available sum rate ${R_{{\rm{sum}}}}$ at the transmitter, then the number of baseband precoding operations per second is expressed as ${\nu _{{\rm{PCD}}}} = \frac{{{R_{{\rm{sum}}}}}}{K\ell}$. At the baseband unit, $\varphi$ can be calculated by $\varphi  = 2{N_{{\rm{RF}}}}K$ \cite{Mohammed14}. Based on (2), the power of linear processing is calculated by
\[\begin{array}{l}
{P_{{\rm{LP\_BB}}}} = \frac{{2\sum\limits_{k = 1}^K {{R_{\rm{k}}}} {N_{{\rm{RF}}}}}}{{\ell {L_{TR}}}}= \\
\frac{{2{N_{{\rm{RF}}}}}}{{\ell {L_{TR}}}}\sum\limits_{k = 1}^K W{\log _2}\left(1 + \frac{{{\bf{h}}^H_{k}{{\bf{B}}_{{\rm{RF}}}}{{\bf{b}}_{{\rm{BB}},{k}}}{\bf{b}}_{{\rm{BB}},{k}}^H{\bf{B}}_{{\rm{RF}}}^H{\bf{h}}_{k}}}{{\sum\limits_{i = 1,i \ne k}^K {{\bf{h}}^H_{k}{{\bf{B}}_{{\rm{RF}}}}{{\bf{b}}_{{\rm{BB}},{i}}}{\bf{b}}_{{\rm{BB}},{i}}^H{\bf{B}}_{{\rm{RF}}}^H{\bf{h}}_{k}}  + \sigma _{\rm{n}}^2}}\right)
\end{array}.\tag{15}\]

Since the precoding of RF transmission system is performed by phase shifters, its power consumption is calculated by
\[{P_{{\rm{LP\_RF}}}} = {N_{{\rm{shifter}}}}{P_{{\rm{shifter}}}},\tag{16}\]
where  ${N_{{\rm{shifter}}}}$ is the number of phase shifters and  ${{P}_{{\rm{shifter}}}}$ is the power of a phase shifter.
Substitute (15) and (16) into (14), the power of linear processing is calculated by
\[{P_{{\rm{LP}}}} = \frac{{2\sum\limits_{k = 1}^K {{R_{k}}} {N_{{\rm{RF}}}}}}{{\ell{L_{{\rm{TR}}}}}} + {N_{\rm{T}}}{P_{{\rm{shifter}}}}.\tag{17}\]

The power to run the precoding algorithm can be calculated by
\[P_{\rm{complex}} = \frac{\Theta C_{\rm cmplx}}{{L_{{\rm{TR}}}}},\tag{18}\]
where $\Theta$ denotes the complexity of the proposed algorithm, $L_{\rm{TR}}$ denotes the transmitter efficiency and $C_{\rm cmplx}$ denotes a constant factor.

Substitute (12), (14), (17) and (18) into (11), and the computation power of multi-user massive MIMO communication systems is derived by
\[\begin{array}{l}
{P_{{\rm{CMPT}}}} = {P_{{\rm{COD}}}}\sum\limits_{k = 1}^K {{R_{k}}}  + \frac{{2\sum\limits_{k = 1}^K {{R_{k}}} {N_{{\rm{RF}}}}}}{{\ell{L_{{\rm{TR}}}}}} + {N_{\rm{T}}}{P_{{\rm{shifter}}}} \\
\begin{array}{*{20}{c}}
{}
\end{array}
+ \frac{\Theta C_{\rm cmplx}}{{L_{{\rm{TR}}}}}
+ K{N_{{\rm{ray}}}}{\kappa ^2}\frac{2}{{\overline \gamma  }}\left( {\frac{{({\kappa ^2} - 1){{\log }_\kappa }N}}{\delta } - 2} \right)\sum\limits_{s = 1}^{{{\log }_\kappa }N} {\frac{1}{{{G^{{\rm{BS}}}}(s)}}}
\end{array}.\tag{19}\]

Furthermore, substituting (19) and (10) into (6), the total power of the transmitter is given by
\[\begin{array}{l}
{P_{{\rm{total}}}} = \frac{1}{\alpha }\sum\limits_{k = 1}^K {{{\left\| {{{\bf{B}}_{{\rm{RF}}}}{{\bf{b}}_{{\rm{BB}},{k}}}} \right\|}^2}}  + {N_{{\rm{RF}}}}{P_{{\rm{one\_RF}}}}\\
\begin{array}{*{20}{c}}
{}&{}&
\end{array}
+ K{N_{{\rm{ray}}}}{\kappa ^2}\frac{2}{{\overline \gamma  }}\left( {\frac{{({\kappa ^2} - 1){{\log }_\kappa }N}}{\delta } - 2} \right)\sum\limits_{s = 1}^{{{\log }_\kappa }N} {\frac{1}{{{G^{{\rm{BS}}}}(s)}}} \\
\begin{array}{*{20}{c}}
{}&{}&
\end{array}
+ {P_{{\rm{COD}}}}\sum\limits_{k = 1}^K {{R_{k}}}  + \frac{{2\sum\limits_{k = 1}^K {{R_{k}}} {N_{{\rm{RF}}}}}}{{\ell{L_{{\rm{TR}}}}}}\\
\begin{array}{*{20}{c}}
{}&{}&
\end{array}
+ {N_{\rm{T}}}{P_{{\rm{shifter}}}} + \frac{\Theta C_{\rm cmplx}}{{L_{{\rm{TR}}}}} + {P_{{\rm{fix}}}}
\end{array}.\tag{20}\]

\section{Problem Formulation}
\label{sec3}
\subsection{Energy Efficiency }
\label{sec3-1}
Considering computation and communication power consumption, next we focus on optimizing the energy efficiency of multi-user massive MIMO communication systems for partially-connected structures by optimizing the hybrid precoding matrices of baseband and RF systems. This optimization problem is formed by
\[\begin{array}{l}
\mathop {{\mathop{\rm maximize}\nolimits} }\limits_{{{\bf{B}}_{{\rm{RF}}}} \in {\mathbb{C}^{{N_{\rm{T}}} \times {N_{{\rm{RF}}}}}},\;\;{{\bf{B}}_{{\rm{BB}}}} \in {\mathbb{C}^{{N_{{\rm{RF}}}} \times K}}} \;{\eta _{{\rm{EE}}}} = \frac{{{R_{{\rm{sum}}}}}}{{{P_{{\rm{total}}}}}},\\
\begin{array}{*{20}{c}}
{}&{}&{}&{}&{}&
\end{array}
{\rm{s}}{\rm{.t}}{\rm{.}}\;\;\;{{\bf{B}}_{{\rm{RF}}}} = {\rm{diag}}\{ {{\bf{m}}_{\rm{1}}},...,{{\bf{m}}_{{N_{{\rm{RF}}}}}}\},\\
\begin{array}{*{20}{c}}
{}&{}&{}&{}&{}&{}&{}&
\end{array}\left\| {{{\bf{B}}_{{\rm{RF}}}}{\bf{B}_{{\rm{BB}}}}} \right\|_{\rm{F}}^2 \le {P_{{\rm{max}}}},
\end{array}\tag{21}\]
where ${\eta _{{\rm{EE}}}}$  is the energy efficiency, and ${P_{\max }}$  is the maximum transmission power. $\left\| {{{\bf{B}}_{{\rm{RF}}}}{\bf{B}_{{\rm{BB}}}}} \right\|_{\rm{F}}^2 \\ \le{P_{{\rm{max}}}}$ is the maximum transmission power constraint. Since RF precoding is performed by phase shifters, only the signal phases change. For the partially-connected structure of RF transmission systems, the element amplitude of complex vector: ${{\bf{m}}_i}$  is fixed as 1. We should point out that when (3) and (20) are substituted into (21), the optimization problem of energy efficiency is a non-concave optimization problem.

\subsection{Cost Efficiency}
\label{sec3-2}
Energy efficiency is an important indicator for service providers. For telecommunication equipment providers, the cost efficiency is another important indicator impacting their design strategies. To evaluate the benefits of the partially-connected structure in RF transmission systems, the cost efficiency of the multi-user massive MIMO communication systems is defined by
\[{\eta _{{\rm{cost}}}} = \frac{{{R_{{\rm{sum}}}}}}{{C_{{\rm{total}}}}},\tag{22}\]
where $C_{{\rm{total}}}$ is the total cost, which is comprised of power consumption cost: $C_{{\rm{power}}}$ and the hardware cost: $C_{{\rm{hardware}}}$ in communication systems. Without loss of generality, the total cost is calculated by
\[C_{{\rm{total}}} = C_{{\rm{hardware}}} + C_{{\rm{power}}},\tag{23}\]
\[C_{{\rm{power}}} = {\beta _{\rm{power}}} {P_{{\rm{total}}}},\tag{24}\]
\[C_{{\rm{hardware}}} = {\beta _{\rm{T}}}{N_{\rm{T}}} + {\beta _{{\rm{shifter}}}}{N_{{\rm{shifter}}}} + {\beta _{{\rm{RF}}}}{N_{{\rm{RF}}}} + {\beta _{{\rm{BB}}}},\tag{25}\]
where  ${\beta _{\rm{power}}}$ is the power rate, ${\beta _{\rm{T}}}$ is the cost coefficient per antenna, ${\beta _{{\rm{shifter}}}}$  is the cost coefficient per phase shifter, ${\beta _{{\rm{RF}}}}$ is the cost coefficient per RF chain and ${\beta _{{\rm{BB}}}}$ is the cost efficient per baseband unit.

\section{Hybrid Precoding Design for the Partially-connected Structure }
\label{sec4}
Taking into consideration the complexity and non-concave properties of the optimization problem in (21), it is difficult to directly solve the baseband and RF precoding matrices. Therefore, we first derive the upper bound on the energy efficiency and then propose a suboptimal solution with joint optimized baseband and RF precoding matrices that can approach the upper bound.

\subsection{Upper Bound of Energy Efficiency}
\label{sec4-1}

To derive the upper bound of energy efficiency, the constraints of energy efficiency optimization in (21) are relaxed. Moreover, to simplify derivations, the product of the baseband precoding matrix ${{\bf{B}}_{{\rm{BB}}}}$  and the RF precoding matrix ${{\bf{B}}_{{\rm{RF}}}}$ is replaced by the fully-digital precoding matrix: ${\bf{B}} \in {\mathbb{C}^{{N_{\rm{T}}} \times {K}}}$, \emph{i.e.}, $\bf{B} = {{\bf{B}}_{{\rm{RF}}}}{\bf{B}_{{\rm{BB}}}}$, where the $k{\rm{th}}$ column of ${\bf{B}}$ is denoted as ${{\bf{b}}_{k}}$ which is the baseband precoding vector for the $k{\rm{th}}$  UE.

\emph{Theorem 1 (Upper bound of energy efficiency)}: When the joint precoding matrix ${\bf{B}}$ is a stationary matrix and the value of ${\bf{B}}$ satisfies the following result:
\[{\pmb{\phi}} _{{k}}^{ - 1}{{\pmb{\psi}} _{{k}}}{{\bf{b}}_{{k}}} - {{\bf{b}}_{{k}}} = 0,\;\;k = 1,2...K,\tag{26}\]
with
\[\begin{array}{l}
{\pmb{\phi}} _{{k}} = \frac{2}{\alpha }\sum\limits_{i = 1}^K {{{\bar R}_{{i}}}} {{\bf{I}}_{{N_{{\rm{Tx}}}}}} +\\
\;\;\;\;\;\;
\frac{2}{{\ln 2}}\overline P W\sum\limits_{i = 1,i \ne k}^K {(\frac{{{\bf{h}}^H_{{i}}{{\bf{b}}_{{i}}}{\bf{b}}_{{i}}^H{\bf{h}}_{{i}}}}{{{{\left( {{\delta _{{i}}}} \right)}^2} + {\delta _{{i}}}{\bf{h}}^H_{{i}}{{\bf{b}}_{{i}}}{\bf{b}}_{{i}}^H{\bf{h}}_{{i}}}} \cdot {\bf{h}}_{{i}}{\bf{h}}^H_{{i}})}
\end{array}
,\tag{27}\]

\[\begin{array}{l}
{{\pmb{\psi}} _{{k}}} =
\{
\overline P  \cdot [\frac{4}{{\ln 2}}W\sum\limits_{i = 1,i \ne k}^K {(\frac{{{\bf{h}}^H_{{i}}{{\bf{b}}_{{i}}}{\bf{b}}_{{i}}^H{\bf{h}}_{{i}}}}{{{{\left( {{\delta _i}} \right)}^2} + {\delta _{{i}}}{\bf{h}}^H_{{i}}{{\bf{b}}_{{i}}}{\bf{b}}_{{i}}^H{\bf{h}}_{{i}}}} \cdot {\bf{h}}_{{i}}{\bf{h}}^H_{{i}})}  +
\\
\;\;\;\;\;\;\;\; \frac{2}{{\ln 2}}\frac{{W{\bf{h}}_{{k}}{\bf{h}}^H_{{k}}}}{{\sum\limits_{j = 1}^K {{\bf{h}}^H_{{k}}{{\bf{b}}_{{j}}}{\bf{b}}_{{j}}^H{\bf{h}}_{{k}}}  + \sigma _{\rm{n}}^{\rm{2}}}}] + {\rm{    }}{\sum\limits_{i = 1}^K {{{\bar R}_{{i}}}}}  \cdot ({P_{{\rm{COD}}}} + \frac{{2{N_{{\rm{RF}}}}}}{{\ell{L_{{\rm{TR}}}}}}) {\cdot} \\
\;\;\;\;\;\;\;\;
[\frac{2}{{\ln 2}}\frac{{W{\bf{h}}_{{k}}{\bf{h}}^H_{{k}}}}{{\sum\limits_{j = 1}^K {{\bf{h}}^H_{{k}}{{\bf{b}}_{{j}}}{\bf{b}}_{{j}}^H{\bf{h}}_{{k}}}  + \sigma _{\rm{n}}^{\rm{2}}}} - \\
\;\;\;\;\;\;\;\;
\frac{2}{{\ln 2}}W\sum\limits_{j = 1,j \ne k}^K {(\frac{{{\bf{h}}^H_{{j}}{{\bf{b}}_{{j}}}{\bf{b}}_{{j}}^H{\bf{h}}_{{j}}}}{{{{\left( {{\delta _{{j}}}} \right)}^2} + {\delta _{{j}}}{\bf{h}}^H_{{j}}{{\bf{b}}_{{j}}}{\bf{b}}_{{j}}^H{\bf{h}}_{{j}}}} \cdot {\bf{h}}_{{j}}{\bf{h}}^H_{{j}})}]
\}
\end{array},\tag{28}\]

\[{\delta _{{i}}} = \sum\limits_{j = 1,j \ne i}^K {{\bf{h}}^H_{{i}}{{\bf{b}}_{{j}}}{\bf{b}}_{{j}}^H{\bf{h}}_{{i}}}  + \sigma _{\rm{n}}^2,\tag{29}\]
the upper bound of energy efficiency is achieved for multi-user massive MIMO communication systems.

\emph{Proof:} When the product of the baseband precoding matrix ${{\bf{B}}_{{\rm{BB}}}}$ and the RF precoding matrix ${{\bf{B}}_{{\rm{RF}}}}$ is replaced by the joint precoding matrix ${\bf{B}} \in {\mathbb{C}^{{N_{\rm{T}}} \times {K}}}$, a relaxed optimization problem is formulated as follows;

\[\mathop {\;{\mathop{\rm maximize}\nolimits} }\limits_{{\bf{B}} \in {\mathbb{C}^{{N_{\rm{T}}} \times {K}}}} \;\overline \eta _{{\rm{EE}}} = \frac{{\overline {{R_{{\rm{sum}}}}} }}{{\overline {{P_{{\rm{total}}}}} }},\tag{30}\]
with
\[\overline {{R_{{\rm{sum}}}}}  = \sum\limits_{k = 1}^K W {\log _2}\left(1 + \frac{{{\bf{h}}^H_{{k}}{{\bf{b}}_{{k}}}{\bf{b}}_{{k}}^H{\bf{h}}_{{k}}}}{{\sum\limits_{i = 1,i \ne k}^K {{\bf{h}}^H_{{k}}{{\bf{b}}_{{i}}}{\bf{b}}_{{i}}^H{\bf{h}}_{{k}}}  + \sigma _{\rm{n}}^2}}\right),\tag{31}\]
\[\begin{array}{l}
\overline {{P_{{\rm{total}}}}}  = \frac{1}{\alpha }{\left\| {\bf{B}} \right\|^2} + {N_{\rm{T}}}{P_{{\rm{shifter}}}} + {N_{{\rm{RF}}}}{P_{{\rm{one\_RF}}}} \\
\begin{array}{*{20}{c}}
{}&{}&
\end{array}
+ K{N_{{\rm{ray}}}}{\kappa ^2}\frac{2}{{\overline \gamma  }}\left( {\frac{{({\kappa ^2} - 1){{\log }_\kappa }N}}{\delta } - 2} \right)\sum\limits_{s = 1}^{{{\log }_\kappa }N} {\frac{1}{{{G^{{\rm{BS}}}}(s)}}}\\
\begin{array}{*{20}{c}}
{}&{}&
\end{array}
+ {P_{{\rm{COD}}}}\sum\limits_{k = 1}^K W {\log _2}\left(1 + \frac{{{\bf{h}}^H_{{k}}{{\bf{b}}_{{k}}}{\bf{b}}_{{k}}^H{\bf{h}}_{{k}}}}{{\sum\limits_{i = 1,i \ne k}^K {{\bf{h}}^H_{{k}}{{\bf{b}}_{{i}}}{\bf{b}}_{{i}}^H{\bf{h}}_{{k}}}  + \sigma _{\rm{n}}^{\rm{2}}}}\right)\\
\begin{array}{*{20}{c}}
{}&{}&
\end{array}
+ \frac{{2{N_{{\rm{RF}}}}}}{{\ell{L_{{\rm{TR}}}}}}\sum\limits_{k = 1}^K W {\log _2}\left(1 + \frac{{{\bf{h}}^H_{{k}}{{\bf{b}}_{{k}}}{\bf{b}}_{{k}}^H{\bf{h}}_{{k}}}}{{\sum\limits_{i = 1,i \ne k}^K {{\bf{h}}^H_{{k}}{{\bf{b}}_{{i}}}{\bf{b}}_{{i}}^H{\bf{h}}_k}  + \sigma _{\rm{n}}^2}}\right) \\
\begin{array}{*{20}{c}}
{}&{}&
\end{array}
+\frac{\Theta C_{\rm cmplx}}{{L_{{\rm{TR}}}}} + {P_{{\rm{fix}}}}.
\end{array}\tag{32}\]

Based on differential calculus, the solution of the partial derivative of ${\bar \eta _{{\rm{EE}}}}$ is zero if the value of ${\bar \eta _{{\rm{EE}}}}$ is an extremum. Therefore, the partial derivative of ${\bar \eta _{{\rm{EE}}}}$ is expressed as

\[\nabla {\bar \eta _{{\rm{EE}}}}({\bf{B}}) = \left[{({\bar \eta _{{\rm{EE}}}})_{{{\bf{b}}_1}}}({\bf{B}}),{({\bar \eta _{{\rm{EE}}}})_{{{\bf{b}}_2}}}({\bf{B}}),...,{({\bar \eta _{{\rm{EE}}}})_{{{\bf{b}}_{{K}}}}}({\bf{B}})\right],\tag{33}\]
with
\[({\bar \eta _{{\rm{EE}}}})_{{{\bf{b}}_{{K}}}}({\bf{B}}) = \frac{{{\psi _k}{{\bf{b}}_k} - {\phi _k}{{\bf{b}}_k}}}{{{{\overline P }^2}}},\;\;k = 1,2...K.\tag{34}\]
Let $\nabla {\bar \eta _{{\rm{EE}}}}({\rm\bf{B}}) = \bf{0}$, then (26), (27), (28) and (29) can be derived.

When the value of $\bf{B}$ satisfies (26), the value of ${\bar \eta _{{\rm{EE}}}}$ is a an extremum point. Denoting superscript $(i)$ as the $i$th iteration, since ${\pmb{\phi}} _{{k}}^{\left( {{{i}}} \right)}$  is a Hermitian symmetric positive matrix, ${\pmb{\phi}} _{{k}}^{\left( {{{i}}} \right)}$ can be extended as ${\pmb{\phi}} _{{k}}^{\left( {{{i}}} \right)} = {\bf{Z}}{{\bf{Z}}^H}$, where ${\bf{Z}}$  is a symmetric positive definite matrix. If ${\bf{X}} = {\left[\pmb{\phi} _{{k}}^{\left( {{{i}}} \right)}\right]^{ - 1}}\pmb{\psi} _{{k}}^{\left( {i} \right)}{\bf{b}}_{{k}}^{\left( {{{i}}} \right)}$, we have the following result:
\[\begin{array}{l}
{\left[{({\bar \eta _{{\rm{EE}}}})_{{\bf{b}}_{{k}}^{\left( {{{i}}} \right)}}}({\bf{B}}^{(i)})\right]^H}\left({\bf{X}} - {\bf{b}}_{{k}}^{\left( {{{i}}} \right)}\right)
 =\\
\begin{array}{*{20}{c}}
{}&{}
\end{array}
\frac{2}{{{{\bar P}^2}}}{\left[{\bf{b}}_{{k}}^{\left( {{{i}}} \right)}\right]^H}{\left({{\bf{Z}}^{ - 1}}{\pmb{\psi}} _{{k}}^{\left( {{{i}}} \right)} - {\pmb{\psi}} _{{k}}^{\left( {{{i}}} \right)}\right)^H}\left({{\bf{Z}}^{ - 1}}{\pmb{\psi}} _{{k}}^{\left( {{{i}}} \right)} - {\pmb{\psi}} _{{k}}^{\left( {{{i}}} \right)}\right){\bf{b}}_{{k}}^{\left( {{{i}}} \right)}.
\end{array}\tag{35}\]

Since the right expression of (35) can be formulated as ${A^T}A$, the left expression of (35) is a Hermitian symmetric positive semidefinite matrix. Hence, ${({\bar \eta _{{\rm{EE}}}})_{{\bf{b}}_{{k}}^{\left( {{{i}}} \right)}}}{({{\bf{B}}^{(i)}})^H}\left({\bf{X}} - {\bf{b}}_{{k}}^{\left( {{{i}}} \right)}\right) \ge 0$ is satisfied for all values of ${\bf{b}}_k^{\left( {{{i}}} \right)}$. By starting from any ${\bf{b}}_k^{\left( i \right)}$ and moving to ${\bf{X}}$, ${\bar \eta _{{\rm{EE}}}}{({{\bf{B}}^{(i)}})}$ is a non-decreasing function. When a fully-digital precoding matrix is configured as a stationary point, the result of \cite{Jiang13} proves that the energy efficiency optimization function of MIMO systems can be converged to an upper bound. When the fully-digital precoding matrix $\bf{b}$ is assumed as a stationary point(based on the result of \cite{Jiang13}), the upper bounds of ${\bar \eta _{{\rm{EE}}}}{({{\bf{B}}^{(i)}})}$ can be computed. Consequently, ${\bar \eta _{{\rm{EE}}}}{({{\bf{B}}^{(i)}})}$ is a convergent function and the upper bound of energy efficiency is achieved for multi-user massive MIMO communication systems.

Algorithm 1 is developed to obtain the optimized fully-digital precoding matrix ${{\bf{B}}^{{\rm{opt}}}}$.

\begin{algorithm*}
\begin{spacing}{1.3}
\setcounter{algorithm}{0}
\renewcommand{\algorithmicrequire}{\textbf{Input:}}
\renewcommand\algorithmicensure {\textbf{Output:} }

\begin{algorithmic}
\caption{Upper bound of hybrid precoding design. }

\REQUIRE $K,{N_{\rm{T}}},{N_{{\rm{RF}}}}$\\
\ENSURE ${{\bf{B}}^{{\rm{opt}}}}$\\
$i=0$, {initialize} ${{\bf{B}}^{{\rm{(0)}}}}$ {with random complex value}\\
\textbf{repeat}\\
\quad\textbf{}{compute} {${\pmb{\phi}_k ^{{(i)}}},{\pmb{\psi}_k ^{{(i)}}}, k = 1...K$} {based on (27)(28)(29)}\\
 \quad\textbf{for} {$\mu  = 0:\varpi :1\begin{array}{*{20}{c}}
{}&{}&{}
\end{array}$} // {$\mu $}  {is the step length}, {$\varpi $} {is the step length interval}.   \\
  \quad\quad \quad \textbf{for} {$k = 1:K$}
\begin{flushleft}
\quad\quad \quad\quad ${\bf{temp}}\_{\bf{b}}_{{k}}^{\left( {{{i + 1}}} \right){{(\mu )}}} = \mu _{}^{}{\left[ {{\pmb{\phi}} _{{k}}^{\left( {{i}} \right)}} \right]^{ - 1}}\left[{\pmb{\psi}} _{{k}}^{\left( {{i}} \right)}{\bf{b}}_{{k}}^{\left( {{i}} \right){{(\mu )}}}\right] + (1 - \mu _{}^{}){\bf{b}}_{{k}}^{\left( {{i}} \right){{(\mu )}}}$
\end{flushleft}
  \quad\quad \quad \textbf{end for} {$k$}\\
 \quad\textbf{end for}  {$\mu $}\\
 \quad{use ${\bf{temp\_b}}_k^{(i + 1)(\mu )}$ as the \emph{k}th column to form matrix ${\bf{temp\_B}}^{(i + 1)(\mu )}$}\\
 \quad{find the highest} {${\overline{\eta}_{\rm{EE}}\left( {\bf{temp}}\_{\bf{B}}_{}^{( {{{i + 1}}}){{(\mu )}}} \right)}$}  {and let} {${\bf{B}}_{}^{\left( {{{i + 1}}} \right)} = {\bf{temp\_B}}_{}^{\left( {{{i + 1}}} \right){{(\mu )}}}$}\\
 \quad{${{i = i + 1}}$};\\
\textbf{until}\\
\quad\quad{${\left\| {{\bf{B}}_{}^{(i + 1)} - {\bf{B}}^{(i)}} \right\|_{\rm{F}}} \le \varepsilon_{1},  \;\;\;\;\;   $ // $\varepsilon_{1}$ is the stopping criterion}\\
{ ${{\bf{B}}^{{\rm{opt}}}} = {\bf{B}}^{(i + 1)}$}
\end{algorithmic}\end{spacing}
\end{algorithm*}

\subsection{Hybrid Precoding Matrix Design}
\label{sec4-2}
When the product of hybrid precoding matrices ${{\bf{B}}_{{\rm{RF}}}}{{\bf{B}}_{{\rm{BB}}}}$ approaches to the optimized fully-digital precoding matrix ${{\bf{B}}^{{\rm{opt}}}}$, the energy efficiency ${\eta _{{\rm{EE}}}}$  will approach the upper bound of energy efficiency in multi-user massive MIMO communication systems. Therefore, the optimized baseband and RF precoding matrices, \emph{i.e.}, ${\bf{B}}_{{\rm{BB}}}^{{\rm{opt}}}$  and ${\bf{B}}_{{\rm{RF}}}^{{\rm{opt}}}$  can be solved by minimizing the Euclidean distance between ${{\bf{B}}_{{\rm{RF}}}}{{\bf{B}}_{{\rm{BB}}}}$ and ${{\bf{B}}^{{\rm{opt}}}}$ \cite{Rusu16,Lin16,Bogale14}, which is formulated by
\[\begin{array}{l}
\mathop {\;{\mathop{\rm minimize}\nolimits} }\limits_{{{\bf{B}}_{{\rm{RF}}}} \in {\mathbb{C}^{{N_{\rm{T}}} \times {N_{{\rm{RF}}}}}},\;\;{{\bf{B}}_{{\rm{BB}}}} \in {\mathbb{C}^{{N_{{\rm{RF}}}} \times K}}} \;{\left\| {{{\bf{B}}^{{\rm{opt}}}} - {{\bf{B}}_{{\rm{RF}}}}{{\bf{B}}_{{\rm{BB}}}}} \right\|_{\rm{F}}},\\
\begin{array}{*{20}{c}}
{}&{}&{}&{}&{}&
\end{array}
s.t.\;\;\;{{\bf{B}}_{{\rm{RF}}}} = {\rm{diag}}\left\{ {\bf{m}_{\rm{1}}},...,{\bf{m}_{{\emph{N}_{{\rm{RF}}}}}}\right\},\\
\begin{array}{*{20}{c}}
{}&{}&{}&{}&{}&{}&{}&
\end{array}
\left\| {{{\bf{B}}_{{\rm{RF}}}}{{\bf{B}}_{{\rm{BB}}}}} \right\|_{\rm{F}}^2 \le {P_{\max }}.
\end{array}\tag{36}\]

To solve the optimized baseband and RF precoding matrices, an alternating minimization method is adopted in this paper \cite{Yu16,Peters09,Ye16}. Based on the principle of alternating minimization and without loss of generality, we first fix the RF precoding matrix ${{\bf{B}}_{{\rm{RF}}}}$  and derive a solution of baseband matrix ${{\bf{B}}_{{\rm{BB}}}}$. In this case, (36) is transferred as
\[\begin{array}{l}
\mathop {\;{\mathop{\rm minimize}\nolimits} }\limits_{{{\bf{B}}_{{\rm{BB}}}} \in {\mathbb{C}^{{N_{{\rm{RF}}}} \times K}}} \;{\left\| {{{\bf{B}}^{{\rm{opt}}}} - {{\bf{B}}_{{\rm{RF}}}}{{\bf{B}}_{{\rm{BB}}}}} \right\|_{\rm{F}}},\\
\;\;\;\;\;\;\;\;\;\;\;\;\;{\rm{s}}{\rm{.t}}{\rm{.}}\;\;\;\;\left\| {{{\bf{B}}_{{\rm{RF}}}}{{\bf{B}}_{{\rm{BB}}}}} \right\|_{\rm{F}}^2 \le {P_{{\rm{max}}}}.
\end{array}\tag{37}\]

Based on the result in \cite{Yu16}, (37) is a nonconvex quadratically constrained quadratic program (QCQP). Let ${{\bf{x}}_{\rm{c}}} = \rm{vec}({{\bf{B}}_{\rm{BB}}})$,  ${\bf{b}}_{\rm{c}}^{{\rm{opt}}} = \rm{vec}({{\bf{B}}^{{\rm{opt}}}})$ and ${\zeta _c} = {\bf{I}} _{K} \otimes {\bf{B}_{{\rm{RF}}}}$, where ${{\bf{x}}_{\rm{c}}}$, ${\bf{b}}_{\rm{c}}^{{\rm{opt}}}$  and ${\zeta _{\rm{c}}}$  are complex vectors, and ${\rm{vec}}()$ denotes vectorization. To transfer (37) into a real QCQP, let  ${t^2} = 1$ and

\[{\bf{x}} = \left[ \begin{array}{l}
{\rm{real}}({{\bf{x}}_{\rm{c}}})\\
{\rm{imag}}({{\bf{x}}_{\rm{c}}})\\
t
\end{array} \right], \tag{38}\]
\[\bf{b}_{}^{{\rm{opt}}} =  \left[ \begin{array}{l}
{\rm{real}}({\bf{b}}_{\rm{c}}^{{\rm{opt}}})\\
{\rm{imag}}({\bf{b}}_{\rm{c}}^{{\rm{opt}}})
\end{array} \right],\tag{39}\]
\[{\zeta _{}} =  \left[ \begin{array}{l}
{\rm{real}}({\zeta _{\rm{c}}})\\
{\rm{imag}}({\zeta _{\rm{c}}})
\end{array} \right].\tag{40}\]
As a consequence, (37) becomes a real QCQP as follows
\[\begin{array}{l}
\mathop {{\mathop{\rm minimize}\nolimits} \;\;}\limits_{{\bf{x}} \in \mathbb{R}{^n}} {{\bf{x}}^H}{\bf T}{\bf{x}},\\
\;\;\;s.t.\;\;{{\bf{x}}_{({\rm{n}})}}^2 = 1,\\
\;\;\;{{\bf{x}}^H}{{\pmb{\Gamma}} _1}{\bf{x}} \le \frac{{{P_{\max }}{N_{{\rm{RF}}}}}}{{{N_{\rm{T}}}}},
\end{array}\tag{41}\]
with
\[{\bf{T}} = \left[ {\begin{array}{*{20}{c}}
{{\zeta ^H}\zeta }&{ - {\zeta ^H}{{\bf{b}}^{{\rm{opt}}}}}\\
{ - {{\bf{b}}^{{\rm{opt}}}}^H\zeta }&{{{\bf{b}}^{{\rm{opt}}}}^H{{\bf{b}}^{{\rm{opt}}}}}
\end{array}} \right],\]
\[n = 2K{N_{{\rm{RF}}}} + 1,\]
where ${\bf{x}}_{({\rm{n}})}$ denotes the $n$th value of ${\bf{x}}$.

Considering ${\bf{x}}^H{\bf{Tx}} = {\rm{Tr}}(\bf{x}\bf{x}^\emph{H})$ and let $\bf{X} = \bf{x}{{\bf{x}}^\emph{H}}$, (41) is simplified as
\[\begin{array}{l}
\mathop {{\mathop{\rm minimize}\nolimits} \;\;{\rm{Tr}}}\limits_{{\bf{X}} \in {\mathbb{R}^{n \times n}}} \bf{(TX)},\\
\;\;\;\;\;\;\;\;\;s.t.\;\;{\rm{Tr}}({\pmb{\Gamma} _2}{\bf{X}}) = 1,\\
\;\;\;\;\;\;\;\;\;\;\;\;\;\;\;\;{\rm{Tr}}({\pmb{\Gamma} _1}{\bf{X}}) \le \frac{{{P_{\max }}{N_{{\rm{RF}}}}}}{{{N_{\rm{T}}}}},\\
\;\;\;\;\;\;\;\;\;\;\;\;\;\;\;\;{\bf{X}} \ge 0,\\
\;\;\;\;\;\;\;\;\;\;\;\;\;\;\;\;{\rm{rank}}({\bf{X}}) = 1,
\end{array}\tag{42}\]
with
\[{\pmb{\Gamma} _1} = \left[ {\begin{array}{*{20}{c}}
{{{\bf{I}}_{{{n - 1}}}}}&{\bf{0}}\\
{\bf{0}}&0
\end{array}} \right],\]
\[{\pmb{\Gamma} _2} = \left[ {\begin{array}{*{20}{c}}
{{{\bf{0}}_{n - 1}}}&{\bf{0}}\\
{\bf{0}}&1
\end{array}} \right].\]

Except for the constraint condition ${\rm{rank}}({\bf{X}}) = 1$, the objective function and other constraint conditions in (42) are convex. To obtain an approximate solution of (42), we first relax the constraint condition ${\rm{rank}}({\bf{X}}) = 1$ \cite{Luo10}. Thus, (42) is transferred into a semidefinite relaxation program (SDR) as follows
\[\begin{array}{l}
\mathop {{\mathop{\rm minimize}\nolimits} \;{\rm{Tr}}}\limits_{{\bf{X}} \in {\mathbb{R}^{n \times n}}} ({\bf{TX}}),\\
\;\;\;\;\;\;\;\;\;s.t.\;\;{\rm{Tr}}({{\pmb{\Gamma}} _2}{\bf{X}}) = 1,\\
\;\;\;\;\;\;\;\;\;\;\;\;\;\;\;\;{\rm{Tr}}({{\pmb{\Gamma}} _1}{\bf{X}}) \le \frac{{{P_{\max }}{N_{{\rm{RF}}}}}}{{{N_{\rm{T}}}}},\\
\;\;\;\;\;\;\;\;\;\;\;\;\;\;\;\;{\bf{X}} \ge 0.
\end{array}\tag{43}\]

When a standard convex algorithm, such as the interior point algorithm \cite{Sidiropoulos06} is performed for (43), an optimized solution ${\bf{X}^{{\rm{opt}}}}$  is solved. Since the constraint condition: ${\rm{rank}}({\bf{X}}) = 1$ is removed in (43), the solution of (43), \emph{i.e.},  ${\rm{Tr}}({\bf{T}{X^{{\rm{opt}}}})}$ is the lower bound of (42).

If ${\bf{X}^{{\rm{opt}}}}$  does not satisfy the constraint condition: ${\rm{rank}}({\bf{X}}) = 1$,  ${{\bf{X}}^{{\rm{opt}}}}$ can not be decomposed as the product of two vectors, \emph{i.e.}, $\bf{x{x^\emph{H}}}$. Consequently, the baseband matrix: ${{\bf{B}}_{\rm{{BB}}}}$  cannot be solved from ${{\bf{X}}^{{\rm{opt}}}}$. To solve this issue, the approximate method in \cite{Sidiropoulos06,Tseng03} is adopted to obtain the vector ${\bf{x}}$. Firstly, ${{\bf{X}}^{{\rm{opt}}}}$  is decomposed as ${{\bf{X}}^{{\rm{opt}}}} = {\bf{U}}\pmb{\sum} {{\bf{U}}^H}$, where every column of $\bf{U}$  is the eigenvector of ${\bf{X}^{{\rm{opt}}}}$. $\pmb{\sum} $  is a diagonal matrix and its diagonal entries are the eigenvalues of ${\bf{X}^{{\rm{opt}}}}$. Secondly, let $\bf{x} = \bf{U}{{\pmb{\sum}} ^{{\rm{\frac{1}{2}}}}}\bf{v}$, where  $\bf{v}$ is a random vector and each element is a complex circularly symmetric uncorrelated Gaussian random variable with zero mean and variance of 1. Therefore, a series of approximate solutions with ${\rm E}[{{{\bf{x{x}}}^H}}{\rm{] = }}{\bf{X}^{{\rm{opt}}}}$  are obtained. To simplify the calculation, we select the random vector  $\bf{x}$ which satisfies $\left\| {{{\bf{B}}_{{\rm{RF}}}}{\bf{B}_{{\rm{BB}}}}} \right\|_{\rm{F}}^2 \le {P_{{\rm{max}}}}$. Based on the selected vector $\bf{x}$, an approximate solution of (37), \emph{i.e.}, the corresponding baseband precoding matrix, is obtained.

Since the RF precoding matrix only changes the signal phase but not the signal amplitude, the power constraint, \emph{i.e.},  $\left\| {{{\bf{B}}_{{\rm{RF}}}}{{\bf{B}}_{{\rm{BB}}}}} \right\|_{\rm{F}}^2 \le {P_{\max }}$ can be ignored in our calculations. In this case, (36) is transferred as
\[\begin{array}{l}
\mathop {{\mathop{\rm minimize}\nolimits} }\limits_{{{\bf{B}}_{{\rm{RF}}}} \in {^{{N_{\rm{T}}} \times {N_{{\rm{RF}}}}}}} \;{\left\| {{{\bf{B}}^{{\rm{opt}}}} - {{\bf{B}}_{{\rm{RF}}}}{{\bf{B}}_{{\rm{BB}}}}} \right\|_{\rm{F}}},\\
\;\;\;\;\;\;\;\;\;s.t.\;\;\;{{\bf{B}}_{{\rm{RF}}}} = {\rm{diag}}\{ {{\bf{m}}_{\rm{1}}},...,{{\bf{m}}_{{N_{{\rm{RF}}}}}}\}.
\end{array}\tag{44}\]

${{\bf{B}}_{{\rm{RF}}}}{{\bf{B}}_{{\rm{BB}}}}$ can be regarded as the $i{\rm{th}}$  column of ${{\bf{B}}_{{\rm{RF}}}}$ phasing on the $i{\rm{th}}$  row of ${{\bf{B}}_{{\rm{BB}}}}$ and then the sum of all results, \emph{i.e.},

\[\begin{array}{l}
{{\bf{B}}_{{\rm{RF}}}}{{\bf{B}}_{{\rm{BB}}}} = \left[{{\bf{b}}_{{\rm{RF}}(:,1)}},...,{{\bf{b}}_{{\rm{RF}}(:,{N_{{\rm{RF}}}})}}\right]\left[
\begin{array}{l}
{{\bf{b}}_{{\rm{BB}}(1,:)}}\\
...\\
{{\bf{b}}_{{\rm{BB}}({N_{{\rm{RF}}}},:)}}
\end{array}
\right]\\
\;\;\;\;\;\;\;\;\;\;\;\;\;\;\;
= \sum\limits_{i = 1}^{{N_{{\rm{RF}}}}} {{\bf{b}_{{{\rm{RF}}}(:,{\emph{i}})}}{\bf{b}_{{{\rm{BB}}}({\emph{i}},:)}}}
\end{array}
.\tag{45}\]

Moreover, we assume that the value range of the element in the $i{\rm{th}}$ column of ${{\bf{B}}_{{\rm{RF}}}}$ to be continuous. Thus, the following result is derived
\[\begin{array}{l}
\mathop {\;{\mathop{\rm minimize}\nolimits} }\limits_{{{\bf{B}}_{{\rm{RF}}}} \in {\mathbb{C}^{{N_{\rm{T}}} \times {N_{{\rm{RF}}}}}}} \;{\left\| {{{\bf{B}}^{{\rm{opt}}}} - {{\bf{B}}_{{\rm{RF}}}}{{\bf{B}}_{{\rm{BB}}}}} \right\|_{\rm{F}}}  \\
\Leftrightarrow \;\;\;\;\;\; \rm{phase}({{\bf{B}}^{{\rm{opt}}}}) = \rm{phase}({{\bf{B}}_{{\rm{RF}}}}{{\bf{B}}_{{\rm{BB}}}})
\end{array}
,\tag{46}\]
where $\rm{phase}({{\bf{B}}^{{\rm{opt}}}})$ is the operation to get phase of each element of matrix ${{\bf{B}}^{{\rm{opt}}}}$. So
\[\begin{array}{l}
{\rm{phase}}({{\bf{B}}_{{\rm{RF(}}}}_{{{i}},{{j}})}) = {\rm{phase}}({{\bf{B}}^{{\rm{opt}}}}_{({{i}},:)}{{\bf{B}}_{{\rm{BB}}}}{_{({{j}},:)}^H})\\
\begin{array}{*{20}{c}}
{}&{}&{}&{}&{}&{}&
\end{array}
,{\rm{ }}1 \le {i} \le {N_{\rm{T}}},{\rm{ }}{{j}} = \left\lceil {{{i}} \cdot \frac{{{N_{{\rm{RF}}}}}}{{{N_{\rm{T}}}}}} \right\rceil
\end{array}
,\tag{47}\]
where $\left\lceil {} \right\rceil$ denotes rounding up to an integer.

Based on (37) and (44), an iteration algorithm of the alternating minimization method is developed by Algorithm 2.

Considering the multi-user massive MIMO communication system with partially-connected structure, an optimized hybrid precoding algorithm to approach the upper bound of the energy efficiency is developed by the oPtimal Hybrid precOding with computation and commuNication powEr (PHONE) algorithm.\\

\vspace{-0.3cm}
\begin{algorithm}[H]
{
\begin{spacing}{1.4}
\setcounter{algorithm}{1}
\renewcommand{\algorithmicrequire}{\textbf{Input:}}
\renewcommand\algorithmicensure {\textbf{Output:} }

\begin{algorithmic}
\caption{Baseband and RF precoding matrices. }

\REQUIRE {${{\bf{B}}^{{\rm{opt}}}}$}\\
\ENSURE {${\bf{B}}_{{\rm{BB}}}^{{\rm{opt}}}$}{,} {${\bf{B}}_{{\rm{RF}}}^{{\rm{opt}}}$\\
$i=0$, initialize {${{\bf{B}}_{{\rm{RF}}}}^{(i)}$}  with random phases }\\
\textbf{repeat}\\
\quad\quad $i=i+1$\\
\quad\quad {fix} {${{\bf{B}}_{{\rm{RF}}}}^{({{i-1}})},$} {calculate} {${{\bf{B}}_{{\rm{BB}}}}^{({{i}})}$}\\
\quad\quad {fix} {${{\bf{B}}_{{\rm{BB}}}}^{({{i}})},$} {calculate} {${{\bf{B}}_{{\rm{RF}}}}^{(i)}$}\\
\textbf{until}\\
\quad\quad ${\left\| {{{\bf{B}}^{{\rm{opt}}}} - {{\bf{B}}_{{\rm{RF}}}}^{(i)}{{\bf{B}}_{{\rm{BB}}}}^{(i)}} \right\|_{\rm{F}} < \varepsilon_{2}}\;\;\;\;\;$ \\
\quad\quad\quad\quad\quad\quad\quad\quad\quad\quad\quad\quad // $\varepsilon_{2}$ is the stopping criterion\\
${\bf{B}}_{{\rm{BB}}}^{{\rm{opt}}} = {{\bf{B}}_{{\rm{BB}}}}^{({{i}})}$, ${\bf{B}}_{{\rm{RF}}}^{{\rm{opt}}} = {{\bf{B}}_{{\rm{RF}}}}^{({{i}})}$ \\

\end{algorithmic}
\end{spacing}
}
\end{algorithm}

\begin{algorithm}[H]{
\setcounter{algorithm}{2}

\renewcommand{\algorithmicrequire}{\textbf{Input:}}
\renewcommand\algorithmicensure {\textbf{Output:} }

\begin{algorithmic}

\caption{oPtimized Hybrid precOding with computation and commuNication powEr (PHONE). }

\REQUIRE {$K,{N_{\rm{T}}},{N_{{\rm{RF}}}}$}\\
\ENSURE {${\bf{B}}_{{\rm{BB}}}^{{\rm{opt}}}$}{,} {${\bf{B}}_{{\rm{RF}}}^{{\rm{opt}}}$}\\
\quad {compute ${{\bf{B}}^{{\rm{opt}}}}$  based on Algorithm1} \\
\quad {compute ${\bf{B}}_{{\rm{BB}}}^{{\rm{opt}}}$  and ${\bf{B}}_{{\rm{RF}}}^{{\rm{opt}}}$  based on Algorithm2  }\\

\end{algorithmic}
}
\end{algorithm}

Based on the computational complexity of matrix calculation and the iterative algorithms in \cite{Luo10} and \cite{Golub96}, the computation complexity of Algorithm 1 is estimated as $O(K^2+N_{Tx}^3)$ floating point operations (flops); the computation complexity of Algorithm 2 is estimated as $O(K^{3.5})$ flops; Combining Algorithms 1 and 2, the computational complexity of the PHONE algorithm, i.e. $\Theta$ is estimated as $O(N_{Tx}^3+K^{3.5})$ flops.

\section{Simulation Results and Discussions}
\label{sec5}

\begin{table}[H]
\setlength{\abovecaptionskip}{0mm}
\centering
\caption{Default Values of Simulation Parameters}
\label{tab:1}
{\scalebox{0.75}
{
\begin{tabular}{c|c|c|c}
\toprule
    \textbf{Parameters } & \textbf{Value} & \textbf{Parameters}&\textbf{Value}\\
    \hline
    $\tau $ & 1  & ${W}$ & 200 kHz \\

    ${L_{{\rm{TR}}}}$ & $12.8 \times {10^{9}}$ flops/Watt &${P_{\rm{one\_RF}}}$&12900 mWatt\\
    $\alpha $ & {0.38} &${\beta _{\rm{T}}}$& 188\\
    ${N_{{\rm{ray}}}}$&{20 } &${\beta _{{\rm{shifter}}}}$&1800\\
    ${P_{{\rm{noise}}}}$&-174 dBm/Hz&${\beta _{{\rm{RF}}}}$&7800\\
   ${P_{{\rm{fix}}}}$&{1 Watt }&${\beta _{{\rm{BB}}}}$&6800\\
    ${P_{\max }}$&{33 dBm}& ${\beta _{\rm{power}}} $&0.9\\
    ${P_{{\rm{COD}}}}$& $0.1 \times {10^{{\rm{ - }}9}}$ {Watt/bit/second} & ${C_{\rm{cmplx}}}$ & 1 \\
    ${P_{{\rm{shifter}}}}$ & 88 $\rm mWatt$  & $\kappa$ & 2\\
    $N$ & 64  & $\delta$ & 0.1\\
    $K$ & 5  &  & \\
\bottomrule
\end{tabular}
}}
\end{table}

\begin{figure*}[!t]
\centering
\includegraphics[width=6in]{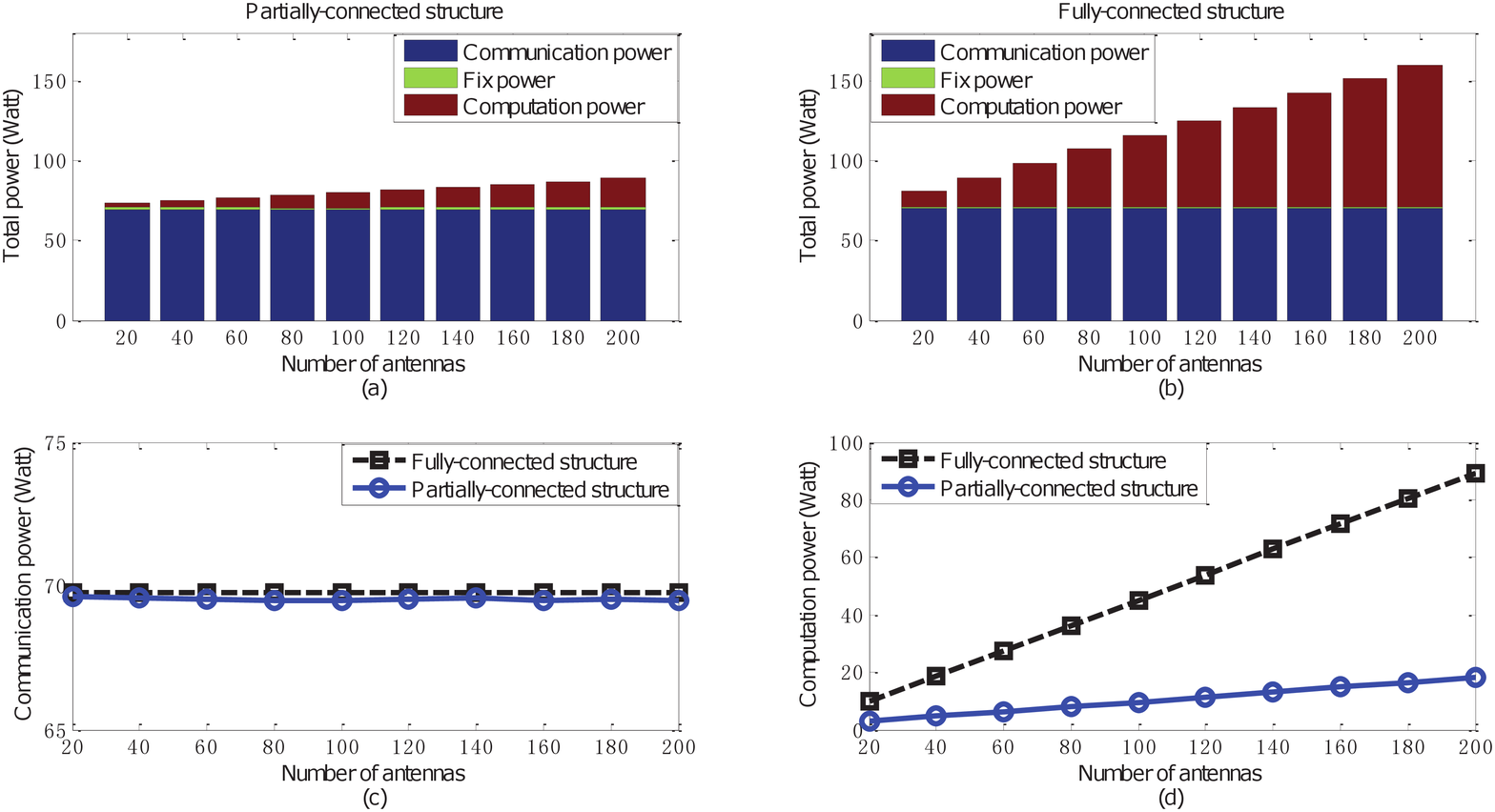}
\begin{quote}
\small Fig. 2. Computation and communication power with respect to the number of transmission antennas considering fully-connected and partially-connected structures.
\end{quote}
\end{figure*}

Energy efficiency optimization solutions with respect to the number of transmission antennas, RF chains, and users are simulated for multi-user massive MIMO communication systems in this section. Without loss of generality, the number of active UEs is configured as 5 and the number of RF chains as 5. Other default values of simulation parameters are listed in Table I. To analyze the proposed PHONE algorithm with a partially-connected structure, the orthogonal matching pursuit (OMP) algorithm \cite{Ayach13} with fully-connected and partially-connected structures are simulated for performance comparisons of multi-user massive MIMO communication systems.

Fig. 2 illustrates the computation and communication power with respect to the number of transmission antennas for fully-connected and partially-connected structures. In this figure, the ``Fully-connected structure" corresponds to the OMP algorithm which is based on the spatial sparse precoding algorithm \cite{Ayach13}, and ``Partially-connected structure" represents the proposed PHONE algorithm. Fig. 2(a) and Fig. 2(b) show that the communication power practically remains flat, whereas the computation power increases. Comparing the results in Fig. 2(a) and Fig. 2(b), the increment of computation power with fully-connected structure is larger than the increment of computation power with partially-connected structure. Moreover, Fig. 2(c) and Fig. 2(d) indicate that the proposed PHONE algorithm with the partially-connected structure outperforms the OMP algorithm with the fully-connected structure due to the saving of communication and computation power in multi-user massive MIMO communication systems.

\begin{figure}
\vspace{0.1in}
\centerline{\includegraphics[width=9.5cm,draft=false]{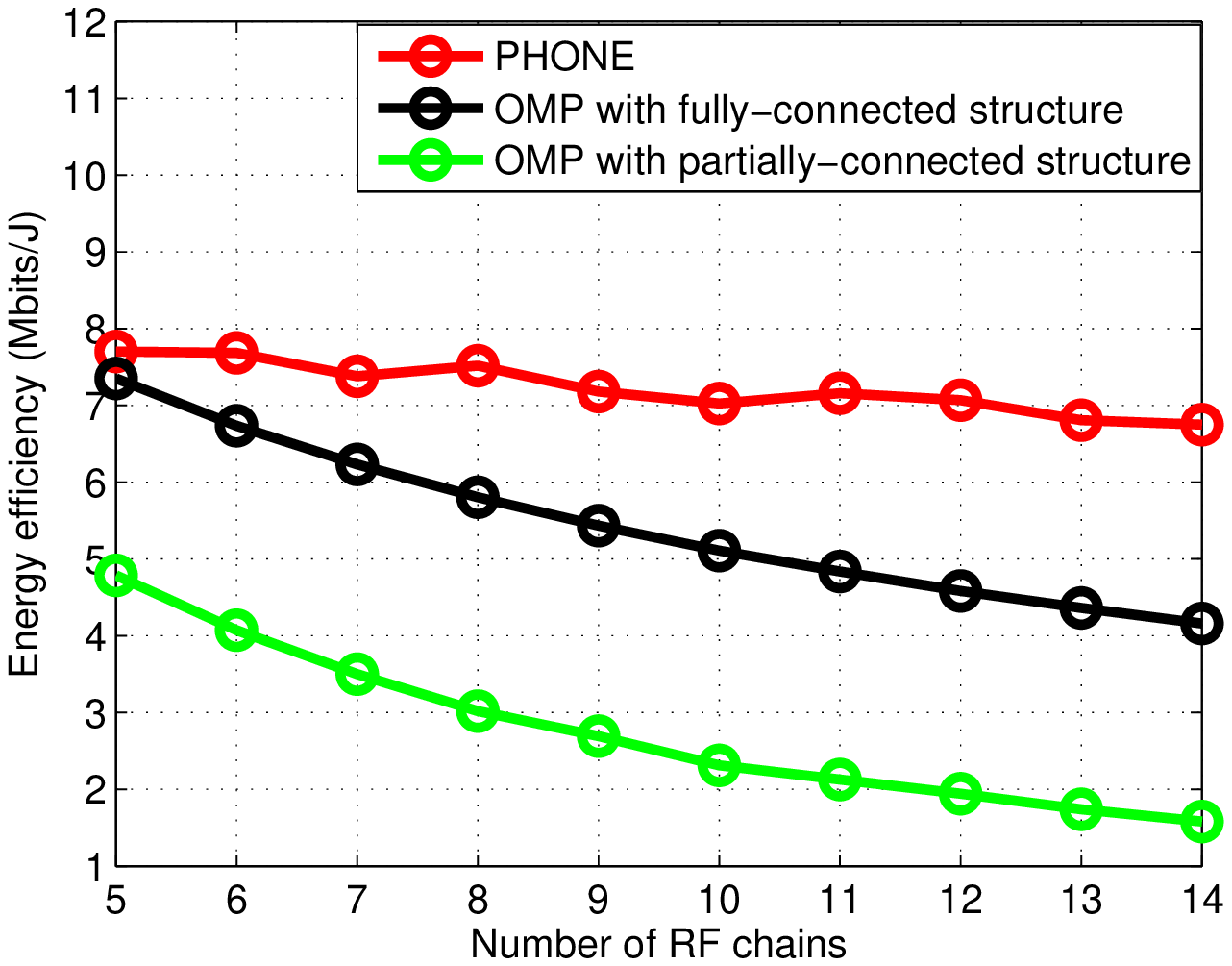}}
{\small Fig. 3(a). \@ \@ Comparing the PHONE and OMP algorithms in terms of energy efficiency with respect to the number of RF chains.}
\end{figure}

Fig. 3(a) depicts the energy efficiency with respect to the number of RF chains. As can be observed, the energy efficiency of the PHONE and OMP algorithms decreases with the increase of the number of RF chains. Moreover, the energy efficiency of the proposed PHONE algorithm is larger compared to the OMP algorithm.

\begin{figure}
\vspace{0.1in}
\centerline{\includegraphics[width=9.5cm,draft=false]{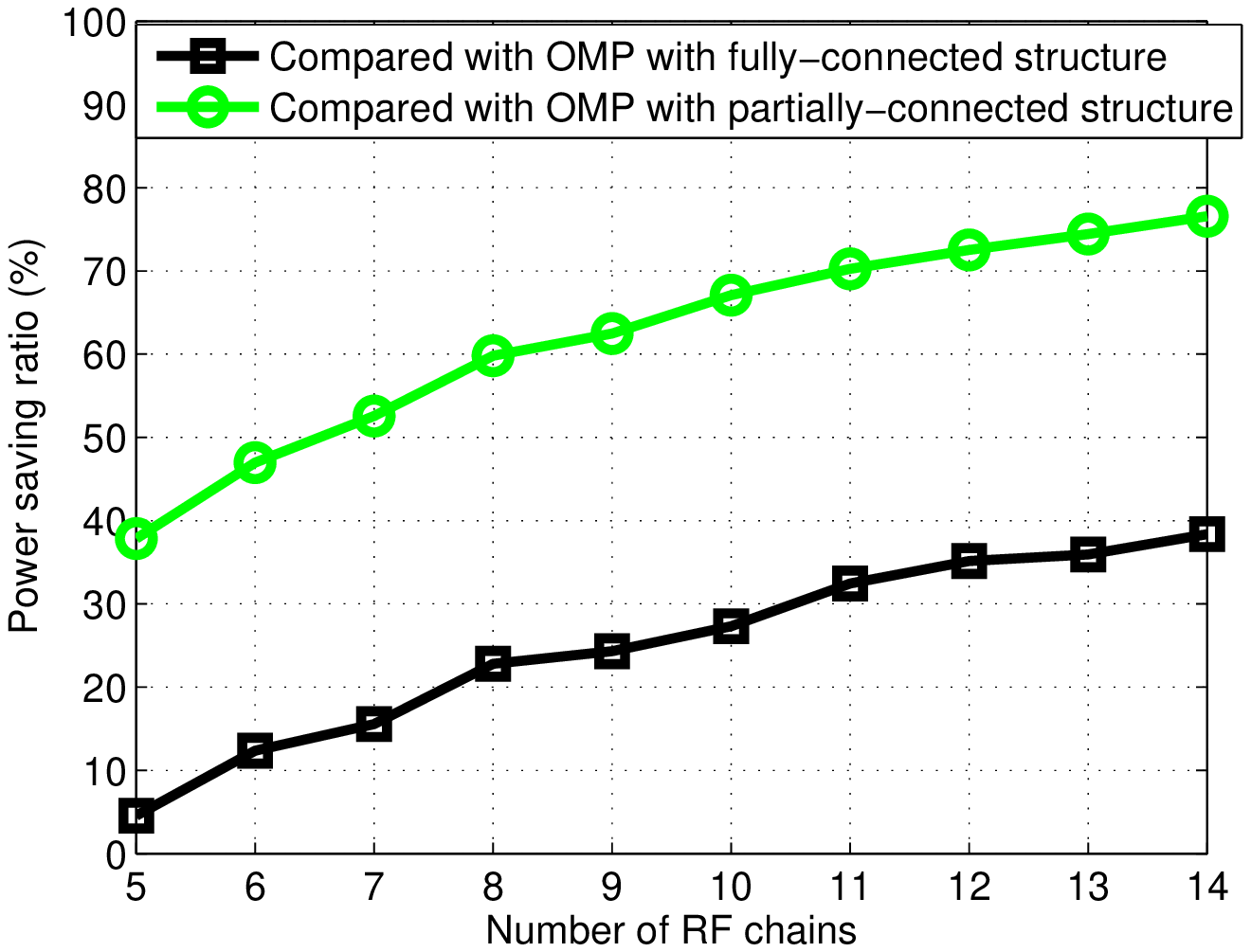}}
{\small Fig. 3(b). \@ \@ Power saving ratio with respect to the number of RF chains.}
\end{figure}

Fig. 3(b) illustrates the power saving ratio with respect to the number of RF chains compared with the OMP algorithm with fully-connected and partially-connected structures in massive MIMO systems. Power saving is defined by the difference of unit rate power consumption. The results in Fig. 3(b) show that the power saving ratio of the PHONE algorithm increases with a greater number of RF chains. For instance, when the number of RF chains is 14, the power saving ratio of massive MIMO system is 76.59\% and 38.38\% compared to the OMP algorithm adopting partially-connected and fully-connected structures, respectively.

\begin{figure}
\vspace{0.1in}
\centerline{\includegraphics[width=9.5cm,draft=false]{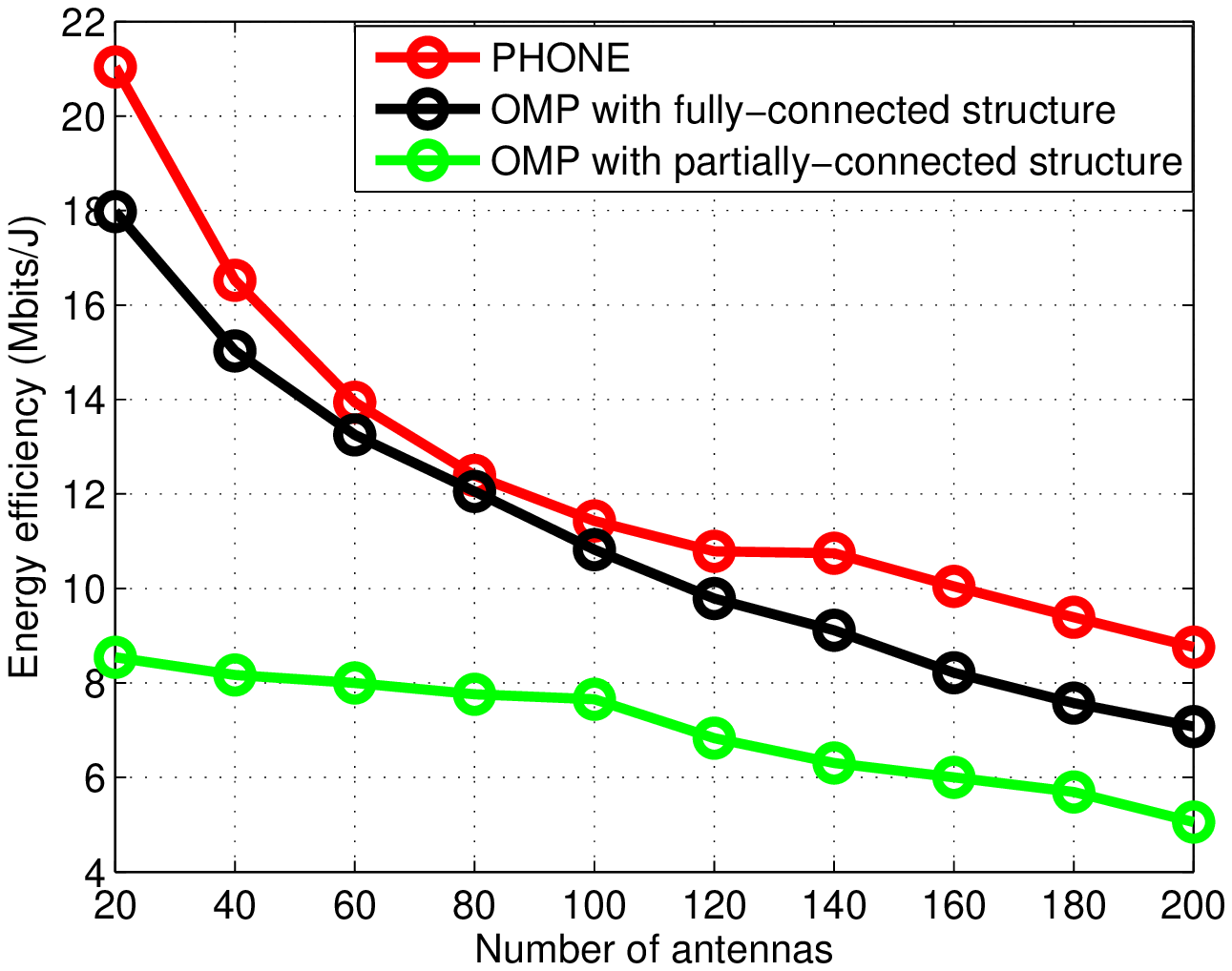}}
{\small Fig. 4. \@ \@ Comparing the PHONE and OMP algorithms in terms of energy efficiency with respect to the number of antennas.}
\end{figure}

Fig. 4 describes the energy efficiency with respect to the number of transmission antennas. When the computation power is considered for massive MIMO systems, the energy efficiency of the PHONE and OMP algorithms decreases with increasing the number of transmission antennas. When the number of transmit antennas is fixed, the energy efficiency of the PHONE algorithm is larger than the energy efficiency of OMP algorithms with fully-connected and partially-connected structures in massive MIMO systems.

\begin{figure}
\vspace{0.1in}
\centerline{\includegraphics[width=9.5cm,draft=false]{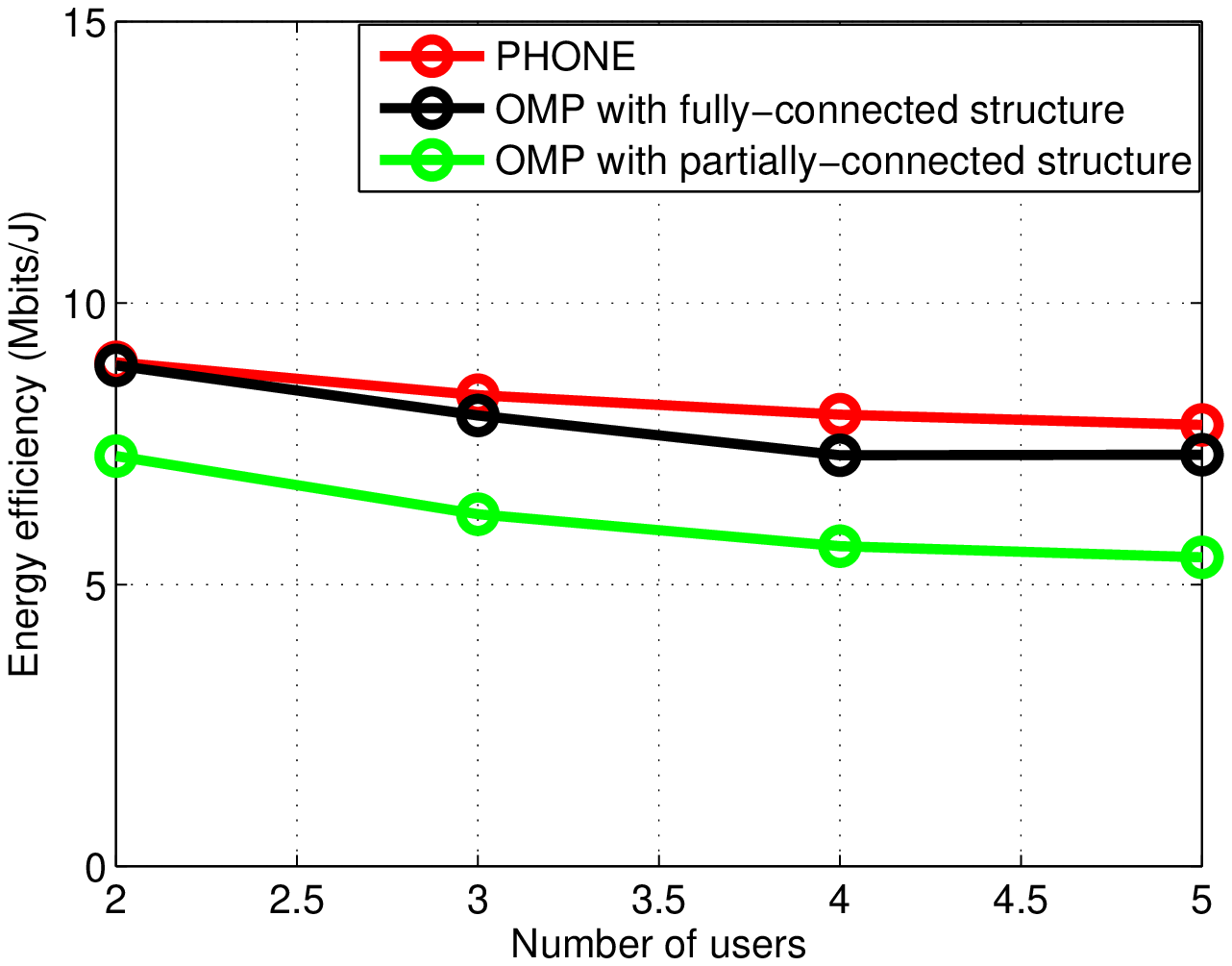}}
{\small Fig. 5. \@ \@ Comparing the PHONE and OMP algorithms in terms of energy efficiency with respect to the number of users.}
\end{figure}

Fig. 5 shows the energy efficiency with respect to the number of users. When the computation power is considered for massive MIMO systems, the energy efficiency of the PHONE and OMP algorithms decreases with increasing the number of users. When the number of users is fixed, the energy efficiency of the PHONE algorithm is larger than the energy efficiency of OMP algorithms with fully-connected and partially-connected structures in massive MIMO systems.

\begin{figure}
\vspace{0.1in}
\centerline{\includegraphics[width=9.5cm,draft=false]{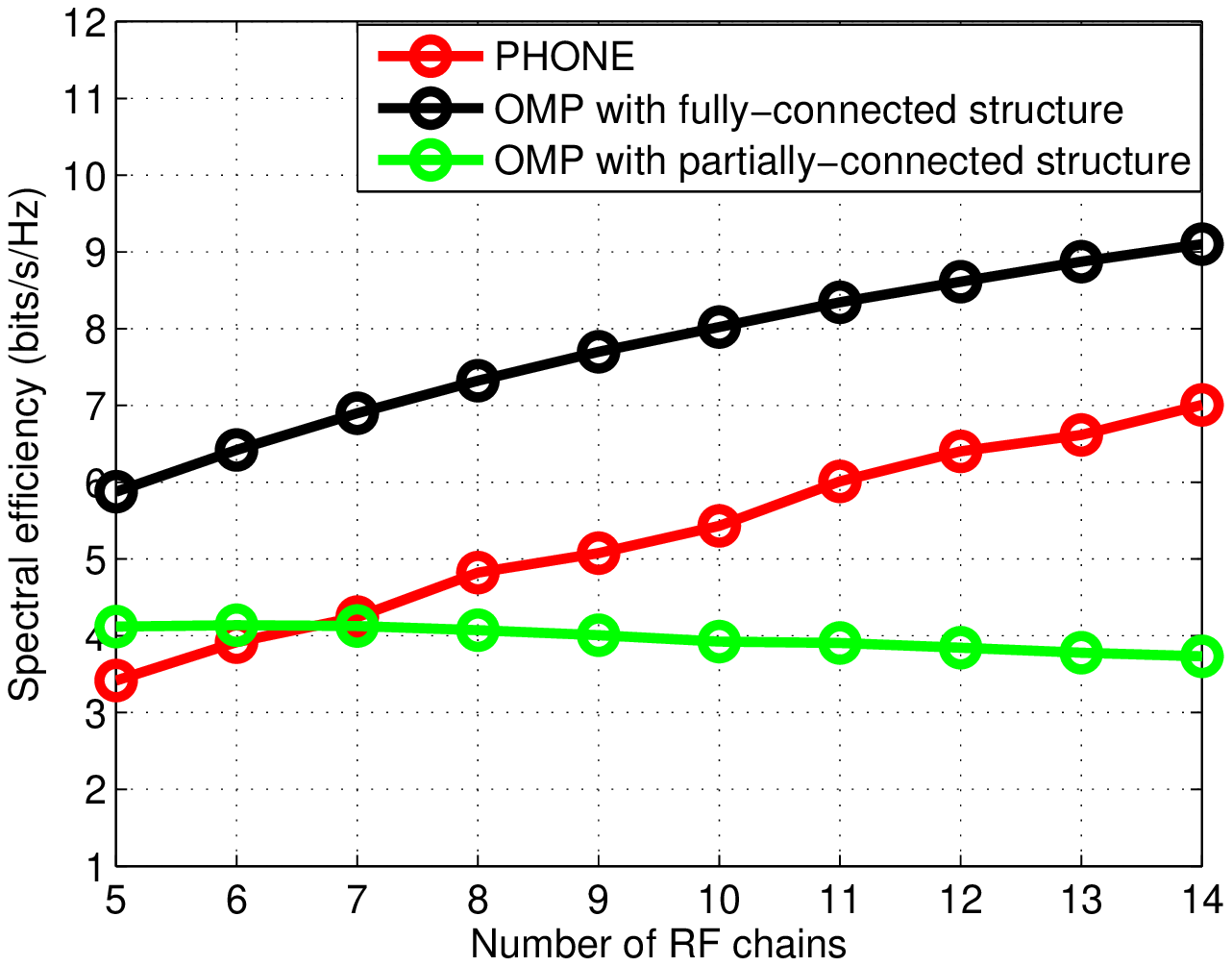}}
{\small Fig. 6. \@ \@ Comparing the PHONE and OMP algorithms in terms of spectral efficiency with respect to the number of RF chains.}
\end{figure}

The spectral efficiency with respect to the number of RF chains is analyzed in Fig. 6. When the computation power is considered for massive MIMO systems, the spectral efficiency of the PHONE algorithm with partially-connected structure and the OMP algorithm with fully-connected structure increases with increasing the number of RF chains. However, the spectral efficiency of the OMP algorithm with partially-connected structure decreases with increasing the number of RF chains. When the number of RF chains is fixed, the spectral efficiency of the PHONE algorithm is less than the spectral efficiency of OMP algorithm with fully-connected structure. When the number of RF chains is less than or equal to 6, the spectral efficiency of the PHONE algorithm is less than the spectral efficiency of OMP algorithm with partially-connected structure. When the number of RF chains is larger than 6, the spectral efficiency of the PHONE algorithm is larger than the spectral efficiency of OMP algorithm with partially-connected structure.

\begin{figure}
\vspace{0.1in}
\centerline{\includegraphics[width=9.5cm,draft=false]{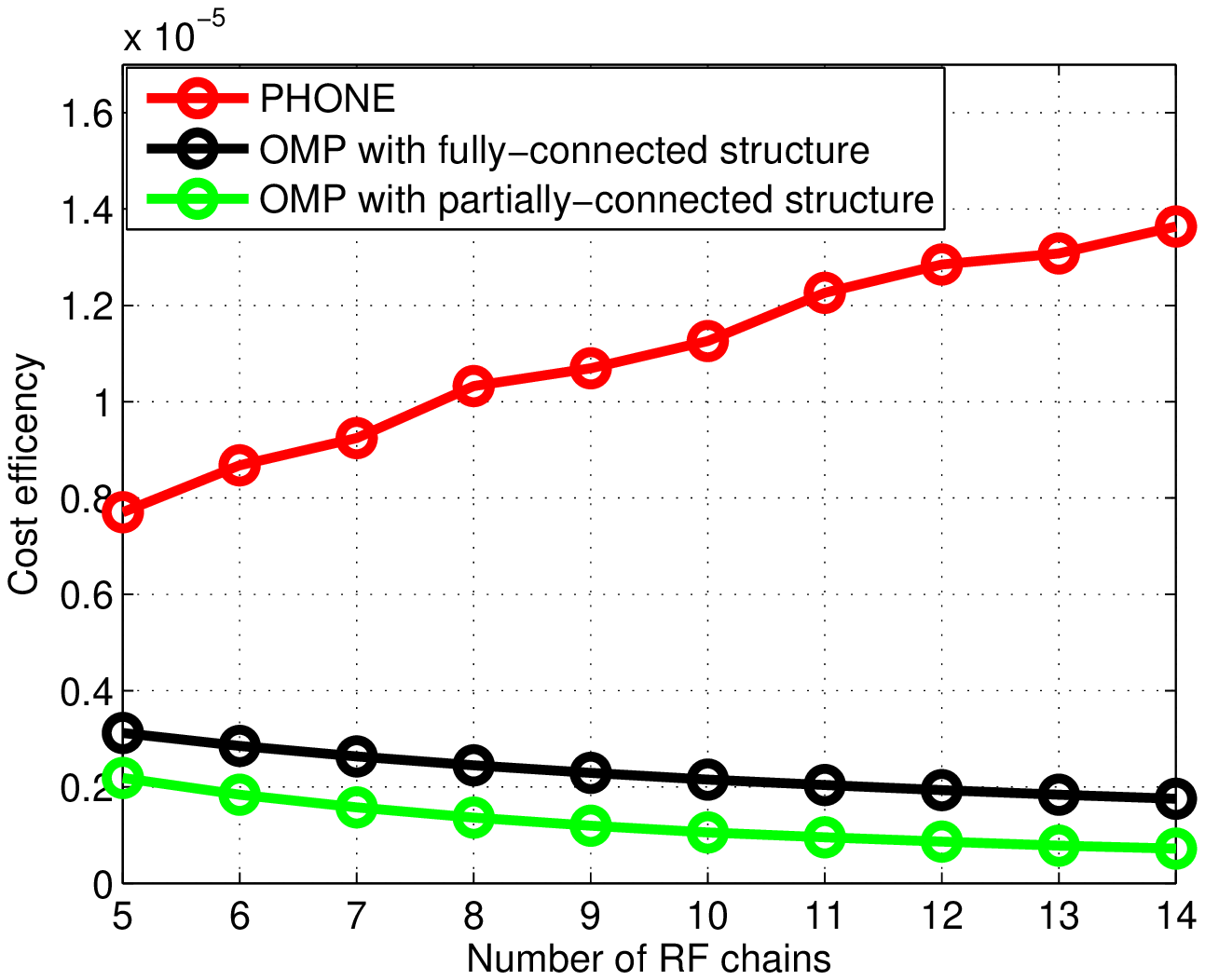}}
{\small Fig. 7. \@ \@ Comparing the PHONE and OMP algorithms in terms of cost efficiency with respect to the number of RF chains.}
\end{figure}

\begin{figure}
\vspace{0.1in}
\centerline{\includegraphics[width=9.5cm,draft=false]{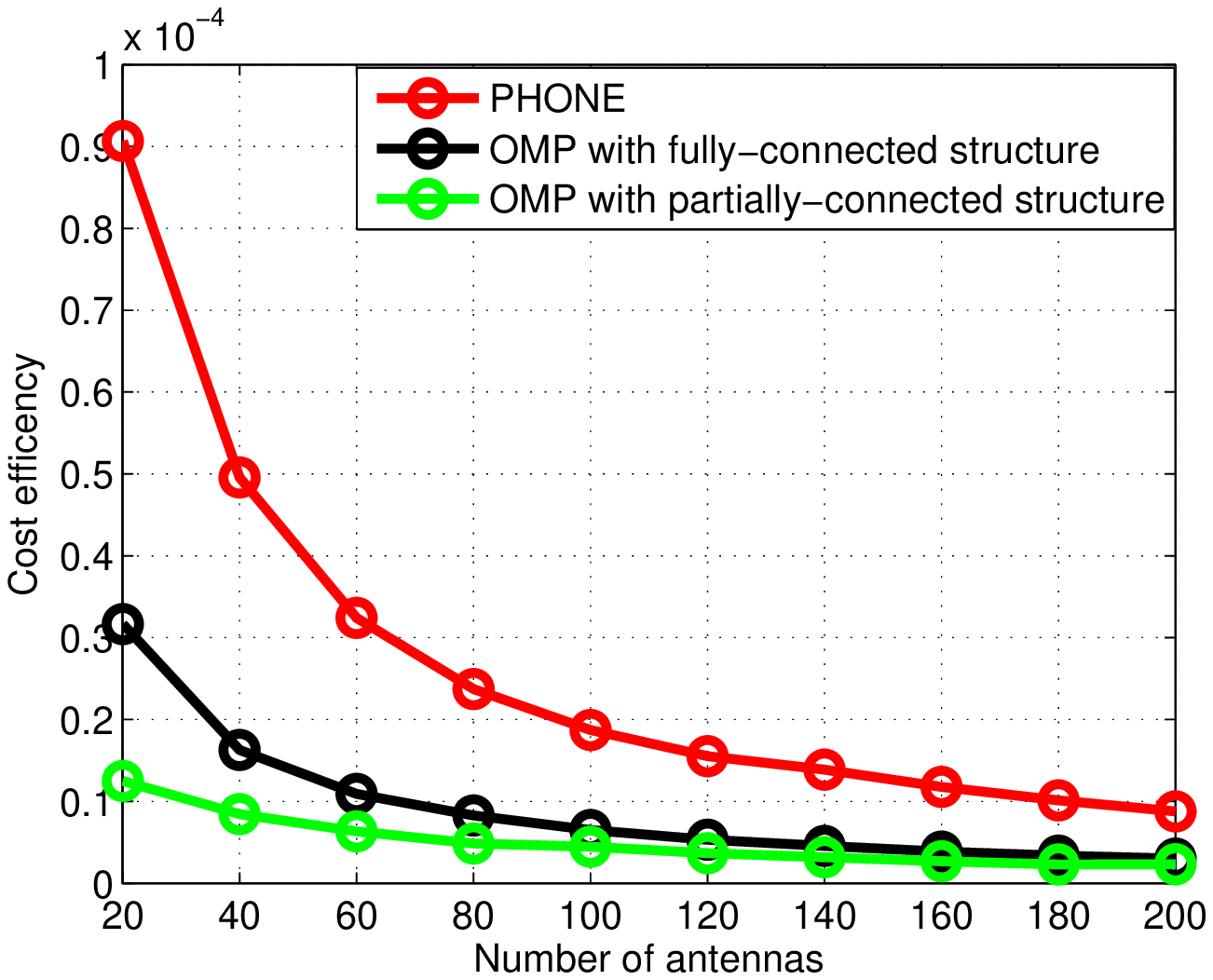}}
{\small Fig. 8. \@ \@ Comparing the PHONE and OMP algorithms in terms of cost efficiency with respect to the number of antennas.}
\end{figure}

The cost efficiency of multi-user massive MIMO communication systems, with respect to the number of RF chains, is compared in Fig. 7. As can be observed, the cost efficiency of the PHONE algorithm improves with the increasing number of transmission antennas; whereas with OMP, it decreases with the increasing number of transmission antennas. When the number of RF chains is fixed, the cost efficiency of the PHONE algorithm outperforms the OMP algorithms with fully and partially connected structures. For instance, compared with the fully and partially connected structures of OMP algorithms, the maximum cost efficiency is improved by 6.78 and 17.97 times, respectively.

The cost efficiency with respect to the number of transmission antennas is compared in Fig. 8. The cost efficiency of multi-user massive MIMO communication systems decreases when the number of transmission antennas for PHONE and OMP algorithms increases. When the number of transmission antennas is fixed, the cost efficiency of the PHONE algorithm is larger than the OMP algorithms with fully-connected and partially-connected structures.

\section{Conclusions}
\label{sec6}
With massive traffic processing, the computation power is emerging as an important part of the energy consumption for 5G massive MIMO communication systems. When computation power is considered, the results of this paper reveal that the energy efficiency of massive MIMO systems decreases when the number of antennas and RF chains increases, which is different than conventional energy efficiency analysis of massive MIMO systems, \emph{i.e.}, only communication power is considered. Faced with this challenge, an optimized solution of energy efficiency that considers communication and computation power is proposed for multi-user massive MIMO communication systems with partially-connected structures. First, an upper bound on energy efficiency of multi-user massive MIMO communication systems is derived. Secondly, the optimized baseband and RF precoding matrices are derived to approach the upper bound on energy efficiency of massive MIMO systems with partially-connected structures. Furthermore, a PHONE algorithm is developed to optimize the performance of multi-user massive MIMO communication system with partially-connected structures. Numerical results indicate that the proposed PHONE algorithm outperforms the OMP algorithm in energy and cost efficiency and the maximum power saving is achieved by 76.59\% and 38.38\% for multi-user massive MIMO communication systems with partially-connected and fully-connected structures, respectively. In future work, we plan to explore the tradeoff between the energy and cost efficiency for multi-user massive MIMO communication systems with partially-connected structures.

\section{Acknowledgement}
\label{sec7}
Dr. Xiaohu Ge would like to acknowledge the
support from China MOST Program of International S\&T Cooperation under grant 2016YFE0133000 and the Hubei
Provincial Science and Technology Department under Grant 2016AHB006.
This research is partially supported by the EU FP7-PEOPLE-IRSES, the
project acronym CROWN under Grant 610524, and EU Horizon 2020 Program under grant 723227, and China International Joint Research Center of Green Communications and Networking under Grant 2015B01008. John Thompson acknowledges part funding of his research by EPSRC grants EP/P000703/1 and EP/L026147/1.

\vspace{-8em}

\begin{IEEEbiography}[{\includegraphics[width=1in,height=1.25in,clip,keepaspectratio]{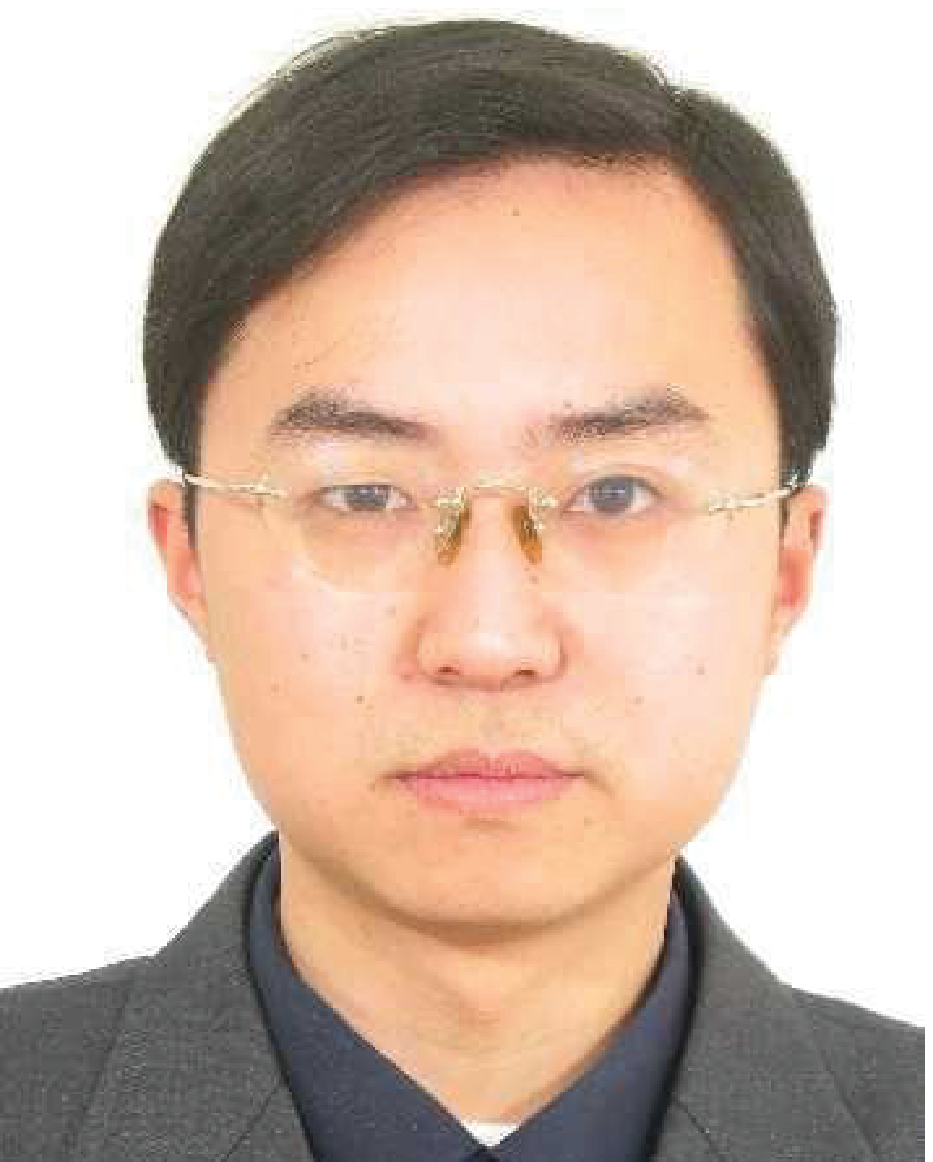}}]{Xiaohu~Ge}
(M'09-SM'11) is currently a full Professor with the School of Electronic Information and Communications at Huazhong University of Science and Technology (HUST), China. He is an adjunct professor with with the Faculty of Engineering and Information
Technology at University of Technology Sydney (UTS), Australia. He received his PhD degree in Communication and Information Engineering from HUST in 2003. He has worked at HUST since Nov. 2005. Prior to that, he worked as a researcher at Ajou University (Korea) and Politecnico Di Torino (Italy) from Jan. 2004 to Oct. 2005. His research interests are in the area of mobile communications, traffic modeling in wireless networks, green communications, and interference modeling in wireless communications. He has published more than 200 papers in refereed journals and conference proceedings and has been granted about 15 patents in China. He received the Best Paper Awards from IEEE Globecom 2010. Dr. Ge served as the general Chair for the 2015 IEEE International Conference on Green Computing and Communications (IEEE GreenCom 2015). He serves as an associate editor for \textit{IEEE Wireless Communications}, \textit{IEEE Transactions on Vehicular Technology} and \textit{IEEE ACCESS}, etc.
\end{IEEEbiography}

\vspace{-5 mm}

\begin{IEEEbiography}[{\includegraphics[width=1in,height=1.25in,clip,keepaspectratio]{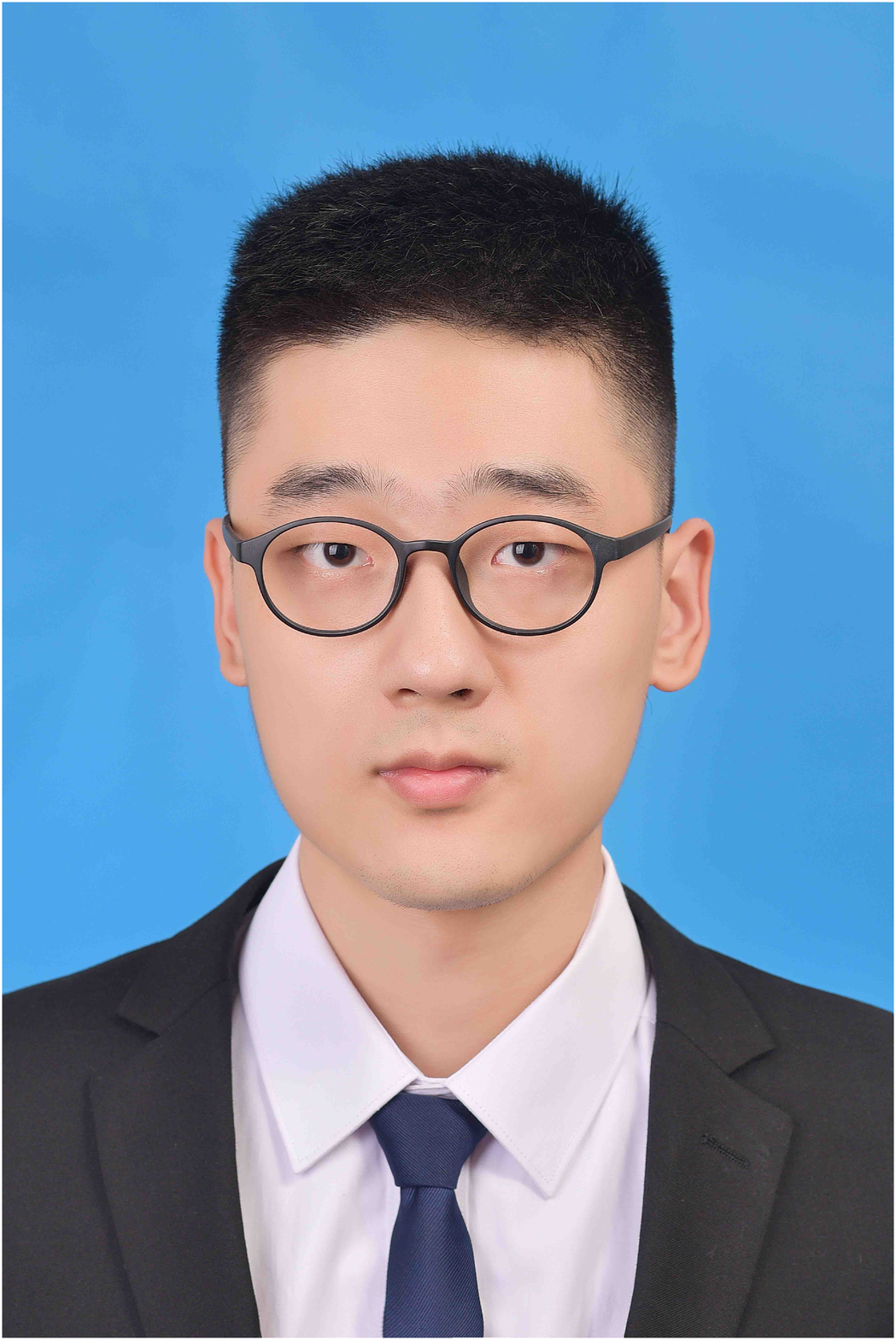}}]{Yang~Sun}
received the B.E. degree with honors in Electronics and Information Engineering and M.S. degree with honors in Information and Communication Engineering from Huazhong University of Science and Technology (HUST), Wuhan, China in 2014 and 2017, respectively. His research interest focus on green communication of massive MIMO systems.
\end{IEEEbiography}

\vspace{-5 mm}

\begin{IEEEbiography}[{\includegraphics[width=1in,height=1.25in,clip,keepaspectratio]{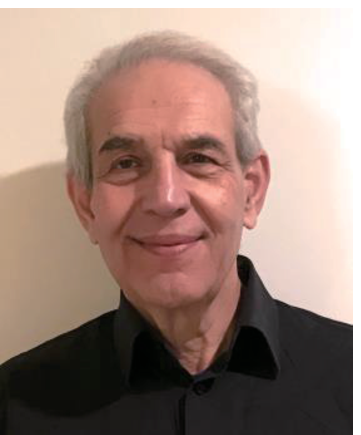}}]{Hamid~Gharavi}
received his Ph.D. degree from Loughborough University, United Kingdom, in 1980. He joined the Visual Communication Research Department at AT\&T Bell Laboratories, Holmdel, New Jersey, in 1982. He was then transferred to Bell Communications Research (Bellcore) after the AT\&T-Bell divestiture, where he became a consultant on video technology and a Distinguished Member of Research Staff. In 1993, he joined Loughborough University as a professor and chair of communication engineering. Since September 1998, he has been with the National Institute of Standards and Technology (NIST), U.S. Department of Commerce, Gaithersburg, Maryland. He was a core member of Study Group XV (Specialist Group on Coding for Visual Telephony) of the International Communications Standardization Body CCITT (ITU-T). His research interests include smart grid, wireless multimedia, mobile communications and wireless systems, mobile ad hoc networks, and visual communications. He holds eight U.S. patents and has over 150 publications related to these topics. He received the Charles Babbage Premium Award from the Institute of Electronics and Radio Engineering in 1986, and the IEEE CAS Society Darlington Best Paper Award in 1989. He was the recipient of the Washington Academy of Science Distinguished Career in Science Award for 2017. He served as a Distinguished Lecturer of the IEEE Communication Society. He has been a Guest Editor for a number of special issues of the Proceedings of the IEEE, including Smart Grid, Sensor Networks \& Applications, Wireless Multimedia Communications, Advanced Automobile Technologies, and Grid Resilience. He was a TPC Co-Chair for IEEE SmartGridComm in 2010 and 2012. He served as a member of the Editorial Board of Proceedings of the IEEE from January 2003 to December 2008. From January 2010 to December 2013 he served as Editor-in-Chief of IEEE Transactions on CAS for Video Technology. He is currently serving as the Editor-in-Chief of IEEE Wireless Communications.
\end{IEEEbiography}

\vspace{-5 mm}

\begin{IEEEbiography}[{\includegraphics[width=1in,height=1.25in,clip,keepaspectratio]{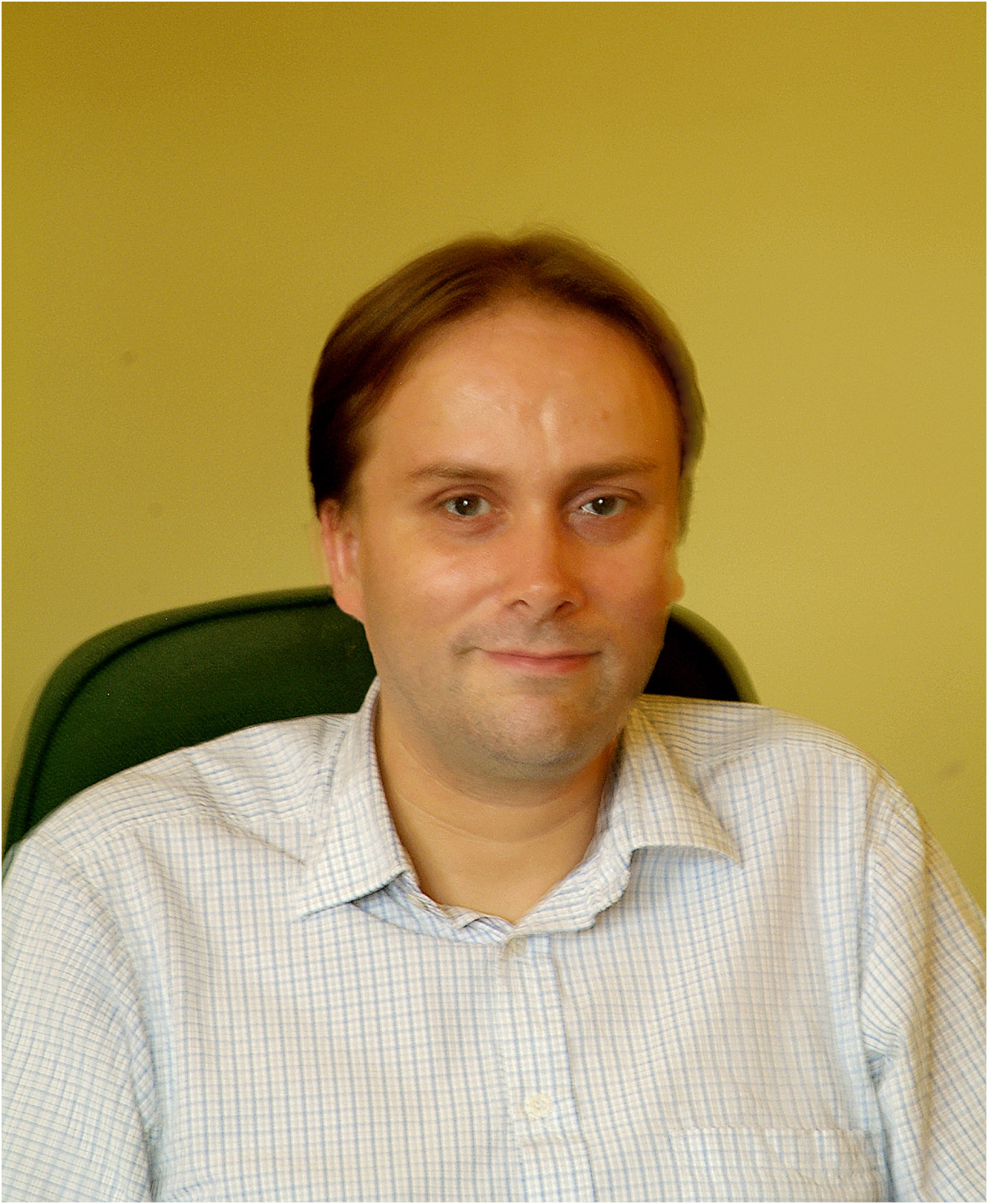}}]{John~Thompson}
is currently a Professor at the School of Engineering in the University of Edinburgh. He specializes in antenna array processing, cooperative communications systems and energy efficient wireless communications. He has published in excess of three hundred papers on these topics, He was coordinator for the recently completed EU Marie Curie Training Network ADVANTAGE, which studies how communications and power engineering can provide future smart grid systems). In 2018, he will be a technical programme co-chair of the IEEE Smartgridcomm conference to be held in Aalborg, Denmark. He currently leads two UK research projects which study new concepts for fifth generation wireless communications. In January 2016, he was elevated to Fellow of the IEEE for contributions to antenna arrays and multi-hop communications. In 2015-2017, he has been recognised by Thomson Reuters as a highly cited researcher.
\end{IEEEbiography}
\end{document}